\documentclass[pre,preprint,showpacs]{revtex4}
\usepackage{doublespace}
\usepackage{graphics}
\usepackage{anysize}
\marginsize{1.5cm}{0.5cm}{0.5cm}{2cm} 
\newcommand{\beq}{\begin{equation}}
\newcommand{\eeq}{\end{equation}}
\newcommand{\barr}{\begin{eqnarray}}
\newcommand{\earr}{\end{eqnarray}}
\newcommand{\bef}{\begin{figure}}
\newcommand{\eef}{\end{figure}}
\newcommand{\svec}[1]{\vec{\mathbf{#1}}}
\begin{document}
\title{Magnetic properties of Ising thin-films with cubic lattices}
\author{Y. Laosiritaworn\dag, J. Poulter\ddag and J.B. Staunton\dag}
\affiliation{\dag\ Department of Physics, University of Warwick,
Coventry CV4 7AL United Kingdom}
\affiliation{\ddag\ Department of Mathematics, Faculty of Science,
Mahidol University, Bangkok 10400, Thailand.}
\begin{abstract}
We have used Monte Carlo simulations to observe the magnetic behaviour of Ising thin-films with
cubic lattice structures as a function of temperature and thickness especially in the critical
region. The fourth order Binder cumulant is used to extract critical temperatures, and an
extension of finite size scaling theory for reduced geometry is derived to calculate the
critical exponents. Magnetisation and magnetic susceptibility per spin in each layer are also
investigated. In addition, mean-field calculations are also performed for comparison. We find
that the magnetic behaviour changes from two dimensional to three dimensional character with
increasing thickness of the film. The crossover of the critical temperature from a two
dimensional to a bulk value is also observed with both the Monte Carlo simulations and the
mean-field analysis. Nevertheless, the simulations have shown that the critical exponents only
vary a little from their two dimensional values. In particular, the results for films with up
to eight layers provide a strong indication of two dimensional universality.
\end{abstract}
\maketitle
%
\section{Introduction}
Studies of the dimensional crossover of magnetic properties from two dimensional ($2d$) to
three dimensional ($3d$) character in magnetic multi-layers have currently attracted much
interest. An understanding of the physical properties of solids as the dimensionality is
reduced is of both technological and fundamental importance \cite{Falicov1990,Johnson1996}. In
particular, studies of ultrathin magnetic films have revealed a number of novel phenomena that
would not have been expected in either $2d$ or $3d$ systems. Of particular interest is the
critical behaviour of magnetic thin-films, for which the dimensionality $d$ is not well
established. It is interesting to consider how magnetic properties such as the magnetisation
$m$, magnetic susceptibility $\chi$ and critical temperature $T_C$ are affected by the
finiteness of the system in the direction perpendicular to the film.

From both theoretical and experimental studies, the critical temperature $T_C$ is found to
increase with film thickness. Nevertheless, how the thickness $l$ of the film relates to the
dimensional crossover of other magnetic critical properties from $2d$ to bulk $3d$ is not fully
understood. Most studies imply that magnetic films belong to a $2d$ universality class
regardless of the thickness. This is commonly understood to be a consequence of the finiteness
of the films along the out-of-plane direction. Close to the phase transition point (critical
temperature $T_C$) the correlation length $\xi$ is constrained by thickness and allowed to
expand only in the in-plane direction. This is certainly a $2d$-like phenomenon. However,
well-known experimental studies of thin-films of nickel \cite{Li1992} have shown contrary
evidence where a dimensional crossover of the critical exponent $\beta$ from $2d$ to $3d$ has
been found. Hence, to clarify this discrepancy, we consider the use of Ising thin-films to study
such a phenomenon in cubic lattices, that is simple cubic (sc), body centred cubic (bcc) and
face centred cubic (fcc).

The Ising model represents a ferromagnet with infinite uniaxial anisotropy. It is useful
because strong magnetic anisotropies are common in ultrathin ferromagnetic films and
monolayers. In addition, it turns out that $2d$ anisotropic Heisenberg systems become
Ising-like near $T_C$. This was shown by Binder and Hohenberg \cite{Binder1994} using Monte
Carlo simulations and has been proven rigorously by Bander and Mills \cite{Bander1988} using
renormalisation group analysis. Also, from the experimental point of view, experiments on
nickel \cite{Li1992} and iron thin-films \cite{Elmers1994} have shown $2d$ results in agreement
with Ising $2d$ critical exponents. These all suggest that the Ising model is a handy tool to
study magnetic thin-films.

Previous studies of Ising thin-films have been by means of the high temperature expansion
\cite{Allen1970,Capehart1976}, renormalisation group \cite{O'Connor1994}, variational cumulant
expansion (VCE) \cite{Lin1992} and Monte Carlo \cite{Binder1974,Schilbe1996} methods. However,
most of them have concentrated on the study of the critical temperature and shift exponent in
the sc lattice. Few have investigated the effective critical exponents in order to determine
how they vary as a function of temperature away from $T_C$. As far as we are aware, nothing
exists in the literature reporting the investigation of critical exponents as a function of
thickness at the critical point. This is mainly due to a presumption that the films are of the
$2d$ class as well as difficulties in the analysis due to finite size effects. Hence, in this
study, we aim to give a more complete picture of the magnetic phase transition in thin-films
especially at the critical point. We have investigated how the magnetic properties vary as a
function of temperature and thickness by means of Monte Carlo simulations and mean-field
theory. We have also calculated the critical temperature and investigated its convergence to
the bulk limit. Most crucially, we have developed an extension to finite size scaling analysis
to observe how the critical exponents vary as a function of thickness. In outline, we firstly
describe the numerical calculations and the methods we have used. Secondly, we show the
evolution of the magnetic properties as a function of temperature and thickness. Then, we
report the critical properties of thin-films in terms of critical temperatures and exponents.
Finally, we discuss our results and compare the characteristic critical exponents with those
found in experiments.

\section{Methodology}
The starting point for the study is the nearest neighbor Ising hamiltonian
   \beq
      H = -J\sum_{<ij>}S_iS_j,
   \eeq
where the spin $S_i$ takes on the values $ \pm 1$ and the sum includes only first
nearest-neighbor (1nn) pairs. Helical and free boundary conditions are used in our simulations
using this hamiltonian for the in-plane and out-of-plane directions respectively. We use units
of $J/k_B$ and $J$ for temperature and energies respectively with the magnetisation per spin
defined as $m = \frac{1}{N}\sum S_i$ where $N$ is total number of spins.

The simulations are carried out for sc, fcc and bcc films of size $N = L \times L \times l$
where $L \times L$ represents the number of sites in each layer of the film and $l$ is the
number of layers in the film i.e. its thickness. We vary $L$ from 64 to 128 (in steps of 8)
with $l$ ranging from a monolayer (bilayer for bcc films) to 20 layers. The spin configurations
of the films are updated using the highly efficient Wolff algorithm \cite{Wolff1989} to
minimise the effect from statistical errors arising from correlation time
\cite{Muller1973,Hohenberg1977}. The random number generator (drand48) is chosen carefully
\cite{L'Ecuyer1998,Entacher1997,Coddington1994} and we ensure that different seed numbers have
no significant effect on the results. For each simulation, roughly $3,000 \times N$ spins are
updated before the system is deemed to have reached equilibrium. From this point, $5\times
10^5$ independent configurations are used to calculate the expectation of magnetisation per
spin $<\!m\!> = \frac{1}{t_{\mathrm{max}}} \sum_i^{t_{\mathrm{max}}} |m_i|$, where
$t_\mathrm{max}$ is total number of measured configurations, and the magnetic susceptibility
$\chi = \beta N(<m^2>-<|m|>^2)$ ( $\beta \equiv J/k_BT$). In a similar fashion, the
layer-dependence of these magnetic properties, $m_k$ and $\chi_k$, where $k$ is a layer index,
are calculated in order to observe the surface effects upon the magnetic properties. For the
investigation of the critical behaviour of the films, we locate the critical temperatures
($T_C$) via the fourth order cumulant $U_L$ \cite{Binder1981} \beq
\label{eqn:fourth_order_cumulant}
      U_L = 1 - \frac{1}{3}\frac{<\!m^4\!>}{<\!m^2\!>^2}.
   \eeq
   At $T=T_C$, $U_L$ should be independent of $L$, i.e. for differing sizes $L$,$L^{\prime}$
 $(U_{L'}/U_L)_{T=T_C} = 1$. Owing  to finite
size effects, we need to plot $T_C(b)$ ($L = bL'$) against $(\ln b)^{-1}$, and extrapolate the
results to the infinite limit i.e. $(\ln b)^{-1} \to 0$ \cite{Binder1981}. To maximise the
efficiency of this $T_C$ calculation, for each thickness, we perform a single long simulation at
a temperature $T_0$ and use the histogram method \cite{Ferrenberg1988,Ferrenberg1991} to
extrapolate $U_L$ to a temperature nearby in order to find the cumulant crossing point. This
temperature is chosen from the temperature at the peak of the magnetic susceptibility curve for
the $L=128$ system, and then around 1 to 4 million spin configurations are used to create the
histograms. To exclude those data obtained from temperatures too far from the simulated
temperature $T_0$, the range of extrapolation $|T-T_0|$ is restricted by the criterion
$|U(T)-U(T_0)| \leq \sigma_E$ \cite{Newman1999} where $U \equiv <\!E\!>$, the average of the
energy, and $\sigma_E$ is a standard deviation of $E$ at $T_0$.

The next step is to extract the critical exponents from our finite size results. To do this we
have developed a finite size scaling method appropriate for films in which  $l/L \ll 1$. The
purpose  is to find how magnetic properties $m$ and $\chi$ scale with the size $L$ and $l$ of
the  systems. The basic finite size scaling  ansatz (e.g.
\cite{Privman1990,Binder1990,Stanley1971,Fisher1971,Barber1983,Fisher1974}) rests on an
assumption that only a  single correlation length $\xi$ is needed to describe the critical
properties of thin-films. As detailed in Appendix \ref{apd:finite_size_scaling} this gives the
following general form for how the magnetisation and susceptibility scale with $L$
   \barr \label{eqn:mchi_scaling_in_thin_films}
      <\!m(T,l)\!> &=& L^{-\beta/\nu}\widetilde{m}(L^{1/\nu}t,l), \nonumber \\
      \chi(T,l) &=& L^{(\gamma/\nu)'}\widetilde{\chi}(L^{1/\nu}t,l),
   \earr
where $(\gamma/\nu)' = \gamma/\nu+2-d$. $\gamma$, $\beta$ and $\nu$ are the critical exponents
associated with $\chi$, $m$ and $\xi$ respectively,  $\widetilde{\chi}$ and $\widetilde{m}$ are
scaling functions for a given $l$ and $t = 1-\frac{T}{T_C}$ is the reduced temperature. These
scaling functions for a range of $L$ should collapse onto a single curve with the correct
critical temperature and exponents. The exponent $1/\nu$ can be extracted from the derivative of
the cumulant with respect to $L$ at $T_C$ owing to its variation with system size as $L^{1/\nu}$
\cite{Binder1981}. Since we have assumed that there is a single $\xi \propto t^{-\nu}$ a single
$\nu$ is to be expected. Note that if equations (\ref{eqn:mchi_scaling_in_thin_films}) correctly
encapsulate the nature of magnetic critical behaviour in films, we can extract the exponents
$\beta/\nu$ and $(\gamma/\nu)'$ by plotting $\log m$ and $\log \chi$ against $\log L$ at $T_C$.
Before this can be done with confidence it is necessary to demonstrate the validity of the
assumption of a single correlation length  and thus the form of equations
(\ref{eqn:mchi_scaling_in_thin_films}). This can be done by establishing the following :-
   \begin{enumerate}
      \item According to eqn. (\ref{eqn:mchi_scaling_in_thin_films}),
      at $T_C$, if we plot $\log \chi \ (m)$ against $\log L$ we should
      get a straight line.
      \item We should find that the sum of $(\gamma/\nu)'+2(\beta/\nu)$ is always equal to 2.
      This again can be derived using the assumption of a single correlation length and
      $\gamma/\nu +2\beta/\nu = d$ \cite{Fisher1969} i.e.
         \beq \label{eqn:hyposcaling_relation_thin_films}
            \left(\frac{\gamma}{\nu}\right)'+2\frac{\beta}{\nu} \quad =
\quad \frac{\gamma}{\nu}+2-d+2\frac{\beta}{\nu}
            \quad = \quad 2,
         \eeq
      \item The clearest demonstration comes from a direct examination of the
      scaling functions themselves, eqn. (\ref{eqn:mchi_scaling_in_thin_films}).
With correct exponents in the critical region, a scaling function for any $L$
but a particular $l$ should collapse onto a single curve. This will also confirm the
reliability of the critical exponents extracted from our simulations
\end{enumerate}

\section{Results and Discussion}
\subsection{Magnetisation and Magnetic Susceptibility}
The magnetisation $m$ and susceptibility $\chi$ profiles for various film thicknesses $l$ and
system sizes $L$ are obtained. The crossover of magnetic behaviour from $2d$-like for the
monolayer (bilayer in bcc) to $3d$-like for films with 20 or more layers is found. For example,
Fig. \ref{figure:mc_ising_sc_films_magnetic_properties_crossover}a,b shows the magnetic phase
transition for a system of $128\times 128\times l$ spins.  The transition point moves from $2d$
to $3d$ values with increasing film thickness.  The layer resolution of these magnetic
properties is also investigated. The influence of the surface on the magnetic properties can
also be seen. For instance, Fig. \ref{figure:mc_ising_sc_films_magnetic_properties_crossover}c
and d show the layer-dependence of $m_k$ and $\chi_k$ for the  $128 \times 128 \times 10$
system. The surface layer has the lowest magnetisation magnitude, whilst the magnetisation per
site increases towards the interior of the films.  Similarly, for $\chi$, representing the
fluctuation of $m$, the strongest magnitude is for the innermost layer and the weakest is at the
surface. The $k^{th}$ and $(l-k)^{th}$ layers have the same magnetic properties owing to the
spatial symmetry and the isotropy of $J$.

The layer variation of the magnetic properties can be explained in terms of finiteness of the
thin-films along the $z$ direction together with the Monte Carlo updating algorithm. In a
simulation run, once a seed spin is chosen in order to select a group of spins to be updated
with the Wolff algorithm, it will have a greater chance of aligning those neighboring spins
which are located towards the interior of the films rather than those towards the surface.
Consider the probability for a site getting updated:
   \beq
      D(i,j,k) = \frac{1}{M}\sum_{n=1}^M s'(i,j,k),
   \eeq
where $s'(i,j,k)$ is the number of times the site at location $(i,j,k)$ is  in a Wolff update,
and $M$ is the total number of clusters forming (flipping) in a simulation. Our simulations show
that, in the ferromagnetic phase, $D$ at the surface layers always has the lowest magnitude,
except, at very low $T$ where the chance of a site getting updated is very high and the same
anywhere. Thus, this implies that once a cluster of spins is formed, the group has its spins
preferentially located in the inner layers rather than near the surface layers. When the group
is flipped, the inner layer will always have the high magnetisation magnitude, but at the
surface layers, there will be more un-flipped spins behaving as a buffer which results in
smaller magnetisation magnitude. Similarly, for the susceptibility, the buffer in the surface
layers helps to reduce the magnetic fluctuation leading to a smaller susceptibility than for the
inner layers.

We find the same qualitative trends by using a mean-field approximation (see appendix
\ref{apd:mean_field}). The evolution of the magnetic properties from $2d$- to $3d$-like is
again found e.g. for sc films in Fig.
\ref{figure:mf_ising_sc_films_magnetic_properties_crossover}a,b. The layer-dependence of
magnetic properties is also evident. As expected, the magnetisation per site is smallest on the
surface layer and highest on the innermost as can be seen, for example, in Fig.
\ref{figure:mf_ising_sc_films_magnetic_properties_crossover}c,d which show results for a
10-layered Ising sc film.

\subsection{Critical Temperatures and Shift Exponents}
We calculate the critical temperatures of the films from our Monte Carlo simulations using the
cumulant method. The results are summarised in Table \ref{table:mc_ising_thin_films}. We find a
change from $2d$ to bulk values as $l$ is increased. Fig. \ref{figure:mc_mf_ising_Tc}a shows
evidence of such a dimensional crossover for the Ising thin-films. Both our $2d$ and bulk
results agree very well with the exact Ising $2d$ results and previous Ising $3d$ studies
\cite{Ferrenberg1991,Adler1983}. As the number of layers is increased, the critical temperature
moves towards the $3d$ value owing to the increase of the average exchange interaction energy.
On similar grounds, for $l \geq 4$, we find $T_C^{\mbox{sc}}(l) < \ T_C^{\mbox{bcc}}(l) < \
T_C^{\mbox{fcc}}(l)$. The mean-field theory calculations show the same qualitative trends (Fig.
\ref{figure:mc_mf_ising_Tc}b) but, as expected, the estimated $T_C$'s are consistently higher
than those from the Monte Carlo work. Moreover an analytic expression for $T_C$ as a function
of film thickness has been made on the basis of a mean-field $T_C$ examination
\cite{Haubenreisser1972}. This is given as,
   \beq
      T_C(l) = T_C(\infty)\frac{z_0+2z_1 \cos(\pi/(l+1))}{z_0+2z_1},
   \eeq
where $z_0$ and $z_1$ are number of nearest neighbors in the same plane and its adjacent plane
respectively. We find our mean-field results to comply with this expression. A summary of
results for $T_C$ from both Monte Carlo and mean-field calculations is presented in Table
\ref{table:mc_ising_thin_films} and Fig. \ref{figure:mc_mf_ising_Tc}.

It is revealing to examine the evolution of the thin-films' critical temperatures from the
monolayer to bulk $3d$ limit in terms of a power law
   \beq \label{eqn:ftting_bulk_Tc_original}
      1-\frac{T_C(l)}{T_C(\infty)} \propto l^{-\lambda},
   \eeq
Here $T_C(\infty)$ is the bulk critical temperature and the shift exponent $\lambda$ has a value
between 1.0 and 2.0 depending on the spin model used and the type of calculation. For thick
films $\lambda$ is expected to be $1/\nu^{3d}$ \cite{Barber1983}. A better fit for films of a
range of thicknesses $l$ is given by \cite{Huang1994} and has the form
   \beq \label{eqn:fitting_bulk_Tc}
      \frac{1}{T_C(l)} =
      \frac{1}{T_C(\infty)}\left[1+\left(\frac{l_0}{l-l'}\right)^{\lambda'}\right],
   \eeq
where $l_0$, $l'$ and $\lambda'$ are all adjustable parameters. Similarly, $\lambda'$ should
tend to $1/\nu^{3d}$ as $l \rightarrow \infty$.

We use eqn. (\ref{eqn:fitting_bulk_Tc}) to fit the $T_C(l)$'s arising from our Monte Carlo
calculations. If it turns out that our $T_C(l)$'s are accurate and that the fit is a useful one,
we should find $T_C(\infty)$ in agreement with the $T_C^{bulk}$ we calculate separately. Results
of the fit are shown in Table \ref{table:mc_mf_fitting_parameters}, and there is less than a 1
percentage difference between $T_C(\infty)$ and $T^{bulk}_C$ for all three types of films. As
can be seen in the Table \ref{table:mc_mf_fitting_parameters}, however, even for $l=20$,
$\lambda'$ is not close to the expected large $l$ value, $1/\nu^{3d}$.  To elucidate further the
evolution from $2d$- to $3d$-like behaviour we rearrange the power law of eqn.
(\ref{eqn:ftting_bulk_Tc_original}) and define
   \beq
      \lambda_{\mathrm{eff}}(l_i) = -\log\left(
      \frac{T_C(\infty)-T_C(l_i)}{T_C(\infty)-T_C(l_{i-1})}\right)/
      \log(\frac{l_i}{l_{i-1}}),
   \eeq
and tabulate $\lambda_{\mathrm{eff}}(l_i)$ with $l_i$ for both our Monte Carlo simulations
($l_i \in \{4,6,8,10,15,20\}$) and mean-field calculations ($l_i$ ranging from 4 to 20 layers)
in Table \ref{table:mc_mf_fitting_parameters}. The $\lambda_{\mathrm{eff}}$'s should converge
to asymptotic shift exponents $1/\nu^{3d}$ when $l$ tends to infinity. A linear least squares
fit between $\lambda_{\mathrm{eff}}(l_i)$ and $1/l_i$ enables us to obtain
$\lambda_{\mathrm{eff}}(\infty)$ which is also given in the Table. (The mean-field values of
$\lambda_{\mathrm{eff}}(\infty)$ for all structures all have values of 2 since the mean-field
$\nu$ is well-known to be $1/2$.) For the Monte Carlo simulations' results, it is gratifying to
find $\lambda_{\mathrm{eff}}(\infty)$ to be close to $1/\nu^{3d}$ which we obtain from our
separate bulk Monte Carlo simulations. This gives good support to the contention of
universality and the asymptotic behaviour contained in eqn.
(\ref{eqn:ftting_bulk_Tc_original},\ref{eqn:fitting_bulk_Tc}).

\subsection{Critical Exponents}
Referring to equations (\ref{eqn:mchi_scaling_in_thin_films}), we notice that the critical
exponents can be extracted from plots of $\log m$ and $\log \chi$, as well as $\log
\frac{\mathrm{d} U_L}{\mathrm{d}\beta}$, against $\log L$ at $T_C$, provided that linear
relations are found. This is actually the first test of the validity of our single correlation
length, $\xi$, assumption. Results from our simulations, at $T_C$, indeed show very good linear
relationships between $\log m$, $\log \chi$ and $\log \frac{\mathrm{d} U_L}{\mathrm{d} \beta}$
with $\log L$ for all our thin-films' thicknesses and structures. We note here that
$\frac{\mathrm{d} U_L}{\mathrm{d} \beta} \propto L^{1/\nu}$. An example is shown in Fig.
\ref{figure:mc_ising_sc_films_exponent_extraction} which presents very good linear fits for
10-layered sc films. Now, the exponents $\beta/\nu$, $(\gamma/\nu)'$, and $1/\nu$ can be
extracted from the slopes of the linear least square fits. They are listed in Table
\ref{table:mc_ising_thin_films}. For the second test, we perform the summation of the computed
exponents $\left(\frac{\gamma}{\nu}\right)'+2\frac{\beta}{\nu}$. We have found that, in each
case, this sum has a value of 2 within error bars as shown in Fig.
\ref{figure:mc_ising_sc_films_exponent}.  According to eqn.
(\ref{eqn:hyposcaling_relation_thin_films}), this passes the second test for the assumption of
a single $\xi$ being sufficient to describe the critical behaviour of the thin-films. For the
last test, we consider the scaling functions in eqn. (\ref{eqn:mchi_scaling_in_thin_films}). We
calculate these scaling functions for the 10-layered sc films with $L = 10, 20, 40, 60, 80,
100, 150$ and $200$ using the exponents extracted from $L = 64$ to $128$. The results are shown
in Fig \ref{figure:mf_ising_sc_scaling_functions}a,b. For large enough $L$ ($L = 60$ to $200$),
excellent data collapses occur. This confirms that our critical exponents are perfectly
reliable. However, in the same figures, for small $L$ (around $L = 10$ and $20$), data
collapsing is not found. This is what we must expect since eqn.
(\ref{eqn:mchi_scaling_in_thin_films}) originated from eqn.
(\ref{eqn:chil_exponent},\ref{eqn:ml_exponent}) (in appendix \ref{apd:finite_size_scaling})
under the condition $l \ll L$. A collapse of scaling functions should not be observed if $L$
approaches $l$. For comparison, we also present the scaling functions for these 10-layered sc
films using the $2d$ monolayer critical exponents in Fig
\ref{figure:mf_ising_sc_scaling_functions}c,d. We can see that the data collapses are not quite
as good. This also confirms the reliability of our results.

Since our results pass all of the three tests above, it can be understood that the single $\xi$
assumption is valid and eqn. (\ref{eqn:mchi_scaling_in_thin_films}) is very useful for
extracting critical exponents from film systems at $T_C$. However, it is unfortunate that we
cannot extract the $\gamma/\nu$ out of $(\gamma/\nu)'$ from our results. This means that we
cannot investigate the dimension $d$ as a function of thickness. Nevertheless, we find that the
exponents $(\gamma/\nu)'$ and $\beta/\nu$, for $l=1$ up to $l=20$ films, are quite close to
their $2d$ values as shown in Table \ref{table:mc_ising_thin_films} and Fig.
\ref{figure:mc_ising_sc_films_exponent}. In particular, for thin-films ($l \leq 8$), the
exponent values seem to be almost identical with the $2d$ results. On the other hand, for
thicker films ($l \geq 10$), a weak dependence of the exponents on $l$ is found. This is
somewhat reasonable since, in the critical region, although the correlation length $\xi$ is
$2d$-like in the sense that it can expand freely only in the $xy$ plane, it is still a function
of thickness. Suppose that at a reduced temperature $t$ close to zero, any film has an in-plane
correlation length $\xi_\parallel$. Then, the total magnitude of the correlation length $\xi
\approx \sqrt{(\xi_\parallel)^2+l^2}$ for different thickness $l$ will also be different. The
thicker the film, the more different from $2d$ and the closer to $3d$ it will be. However,
thick films are beyond the scope of this study and we find that, for thin-films with thickness
around 8 layers and thinner, the exponents are very close to the $2d$ values. Thus, we feel
that it is quite safe to conclude that thin-films belong to the $2d$ class. In addition, our
results show that, at a good level of agreement, the critical exponents from all structures
represent the same universality class.

Finally, we compare our results for the exponent $\beta$ with those available from an experiment
with Ni films \cite{Li1992}. As shown in Fig. \ref{figure:mc_ising_beta_comparison}, the
experimental results are close to ours only for films thinner than 4 layers. This is not
surprising because the anisotropic Heisenberg magnet becomes Ising-like only for very
thin-films. However, instead of a sharp dimensional crossover of $\beta$ at around 5 to 8 layers
\cite{Li1992}, we found a trend which suggests that a dimensional crossover from $2d$- to
$3d$-Ising should occur for much thicker films. This is perhaps indicated by the slight increase
of $\beta/\nu$ towards its bulk value from $l = 10$ onwards. This issue may also be partly
answered in that the dimensional crossover of $\beta$ in \cite{Li1992} may really be a
transition from $2d$-Ising to $3d$-Heisenberg instead of from $2d$- to $3d$-Ising in our study.
So, a direct comparison may not be allowed. Nevertheless, although Ising and Heisenberg films
may be quantitatively different, they should qualitatively share same characteristics since, in
the critical region, the divergence of $\xi$ in both models is $2d$-like. So, it is strange that
$\beta$ in \cite{Li1992} should change its value so abruptly with very few layers. To find more
answers, we consider \cite{Schilbe1996} (Schilbe et al,1996) which claimed `the author of
\cite{Li1992} neglected the dependence of the (effective) critical exponents away from the
critical point on the reduced temperature in their evaluation of the experimental data'. In
other words, the range of temperatures $10^{-3} < t < 10^{-1}$ used for the power law fit in
\cite{Li1992} may not belong to the asymptotic behaviour (critical region), and the fit may lead
to somewhat dubious results. So, to investigate this closely, we follow \cite{Schilbe1996} who
defined
   \beq \label{eqn:beta_effective}
       \beta_{\mathrm{eff}} = \frac{\partial \log m}{\partial \log t},
   \eeq
and study how this varies with $t$ in thin-films.

From Monte Carlo simulations in sc Ising thin-films for temperatures below but close to $T_C$,
results of $\log m$ against $\log t$ for $L\times L \times 10$ sc films are shown in Fig.
\ref{figure:mc_ising_logm_vs_logt}. The figure shows how, close to $T_C$, for each $L$, the
correlation length $\xi$ notices the finite size $L$, that is where the finite size effect
becomes important. Roughly, depending on the value of $\nu$ (see Table
\ref{table:mc_ising_thin_films}), $\xi \propto t^{-\nu}$ will realise its finite limit at a
temperature around $\log t = -\frac{1}{\nu}\log L$. We have also plotted $\beta_{\mathrm{eff}}$
against $\log t$ for $100 \times 100 \times l$ films in Fig. \ref{figure:mc_ising_sc_beta_eff}.
It is found that the pattern of the $\beta_{\mathrm{eff}}$ curves is in agreement with the
rough investigations in \cite{Schilbe1996}. Since $L = 100$ is used in our calculation, the
$\beta_{\mathrm{eff}}$ are reliable (with minimum finite size effect) only down to $\log t
\simeq -2$. However, this range of $\log t$ does cover most of the range $-3 < \log t < -1$
used in the power law fit in \cite{Li1992}, and shows a quite interesting phenomenon. It
appears that this range of temperature is not in the critical region because
$\beta_\mathrm{eff}$ for each film is not constant, but peaks at a certain temperature. Outside
the critical region, the correlation length grows as we increase the temperature towards $T_C$.
As long as $\xi$ is smaller than the film thickness, the film tries to behave as the $3d$ bulk
system and $\beta$ grows somewhat towards the bulk value $\beta^{3d} \approx 0.3258$
\cite{Ferrenberg1991}. However, at some temperature $\xi$ becomes comparable with the film
thickness and is only allowed to expand in the in-plane direction, that is $2d$ behaviour.
Thus, $\beta$ changes its trend and the resulting decrease results in a peak. We can also
notice that the thicker the film, the closer the peak can grow towards the bulk value. Of
course, if one tries to perform the power law fit in this range of temperature, a sudden change
of $\beta_{\mathrm{eff}}$ will be observed.

We may conclude that the range $-3 < \log t < -1$ used for the power law fit
in \cite{Li1992} is outside the critical region and the power law fit should
not work well. On the other hand, it must be emphasised that our exponents
are extracted at the critical temperature by means of finite size scaling. Our
exponents are the real critical exponents. This explains why the behaviour
of $\beta$ in \cite{Li1992} should not relate to our results.

\section{Conclusion}
We have studied  the magnetic behaviour of Ising thin-films in sc, bcc, and fcc structures
using extensive Monte Carlo simulations. We have found the dimensional crossover of both the
magnetisation $m$ and the magnetic susceptibility $\chi$ from $2d$- to $3d$-like with
increasing film thickness. The layer-components  of $m$ and $\chi$ have also been presented.
The surface layers are found to have the lowest magnitudes whilst the innermost layers have the
highest, a trend which is derived from the free boundary at the surface. (This behaviour is
also found in our mean-field study.) We track how the films' critical temperatures, $T_C$,
evolve from $2d$ to $3d$ values with increasing film thickness. Results for square monolayers
agree well with the exact Ising $2d$ $T_C$ and our bulk $3d$ results are also in good agreement
with earlier studies in the literature. We are also able to fit our calculated $T_C$'s for
films of varying thickness, $l$, involving 3 parameters (shift exponents). This expression
could be examined in the limit of infinitely thick films and the limit agrees very well with
that calculated directly from $3d$ simulations. This fitting expression could also be used on
the mean-field calculations.

We examine the critical regime of these systems in detail and extract critical exponents. For
this purpose we develop a finite size scaling method for films. It is based on one assumption
that a single correlation length $\xi = \xi(l)$ is sufficient to describe the critical
behaviour of Ising films. The validity of this assumption is successfully verified. From our
results we find a very weak variation of the critical exponents with respect to $l$, the film
thickness. For thin-films ($l \leq 8$) the exponents are essentially the same as $2d$ and from
this it can be implied that thin-films fall into the $2d$ class. For thicker films, however,
($l \geq 10$) a weak $l$ dependence is noticeable because $l$ is thick enough for the
correlation length to distinguish the geometry of the films from that of a simple $2d$ lattice.
As an example we present figures of the scaling functions for 10-layered sc films which are
constructed by using both their own 10-layer critical exponents and also the well-known $2d$
exponents. For large enough $L$ (layer extent) we find good collapsing ($L$-independence) for
the thickness-specific exponents which underscores the accuracy of our results. This collapse
degenerates for $L \rightarrow l$ which confirms that our finite size scaling analysis is valid
only for $L \gg l$. When we use the $2d$ critical exponents in the scaling functions a slightly
poorer data collapse is found which points to the real disparity between thin-films and $2d$
exponents.

In comparison with the experimental results described in \cite{Li1992} our results for the
thinner films bear up well. A direct comparison, however, is not possible. This is because the
experimental data from nickel films on a tungsten substrate can be interpreted to show a
transition from $2d$-Ising to $3d$-Heisenberg rather than from $2d$-Ising to $3d$-Ising which is
the only possibility for the model we have studied here. There is also the issue that the
experimental data used to make a power law fit are taken from temperatures that are outside the
critical regime.

{\it Acknowledgements} We acknowledge the use of computer resources provided by the Centre for
Scientific Computing at the University of Warwick. One of us (Y. Laosiritaworn) would like to
thank The Institute for The Promotion of Teaching Science and Technology (Thailand) for
financial support (under DPST project) on his study.

\appendix
\section{Finite Size Scaling} \label{apd:finite_size_scaling}
We assume that the films' extent shows up when the correlation length growing along the $z$
direction ($\xi_z$) is about the same size as $l$, but $\xi_x$ and $\xi_y$, the in-plane $x$ and
$y$ correlation lengths, are still smaller than $L$ and approach $L$ when $T \to T_C$. So, for a
film in the critical region, we propose a hypothesis that a single $\xi$ is enough to describe
the critical behaviour i.e. $\xi_x = \xi_y = \xi$ with $\xi_z = l$ as a constant. This implies
$\xi = \xi(l)$. From this we modify a technique described by Binder and Wang \cite{Binder1989}.
The aim is to find out how thermodynamical properties such as $m$ and $\chi$ scale with $L$ in a
finite size lattice in the critical region where the film thickness $l$ is a constant.

Consider the susceptibility
   \beq
      k_BT\chi =
      \frac{1}{L_xL_yL_z}\sum_{x_1=1}^{L_x}\sum_{y_1=1}^{L_y}\sum_{z_1=1}^{L_z}
      \sum_{x_2=1}^{L_x}\sum_{y_2=1}^{L_y}\sum_{z_2=1}^{L_z}
      <\!m(x_1,y_1,z_1)m(x_2,y_2,z_2)\!>-<\!m\!>^2,
   \eeq
where we set the lattice spacing as 1. We use  periodic boundary conditions along the
$x$ and $y$ directions.
Changing the index $z_2$ to $z_1+z$ under a condition that $L_x$ and $L_y \gg 1$, we can
write
   \beq
      k_BT\chi \simeq
      \frac{1}{L_z}\int_0^{L_x}\mathrm{d}x\int_0^{L_y}\mathrm{d}y\sum_{z_1=1}^{L_z}
      \sum_{z=1-z_1}^{L_z-z_1}
      <\!m(0,0,z_1)m(x,y,z_1+z)\!>-<\!m\!>^2.
   \eeq
Next, suppose that the two-point correlation function for an isotropically shaped system at
$T_C$ takes the form \cite{Kadanoff1966}
   \beq \label{eqn:g2c}
      G^{(2)}_c(r,T_C)\ \equiv\ \left.<\!m(0)m(r)\!>-<\!m\!>^2\right|_{T_C}\
                       \sim\ \frac{1}{r^{d-2+\eta}}\ \sim\ (x^2+y^2+z^2)^{-(d-2+\eta)/2}
   \eeq
(where $r = |\svec{r}_1-\svec{r}_2|$, $d$ is the dimension of the system, and $\eta$ is a
critical exponent) and that this form is also valid for anisotropically shaped systems like
films, we rewrite the susceptibility
   \beq
   k_BT_C\chi(T_C) \sim \frac{(L_xL_y)}{L_z}^{1-(d-2
    +\eta)/2}\int_0^1\mathrm{d}x'\int_0^1\mathrm{d}y'
         \sum_{z_1=1}^{L_z}\sum_{z=1-z_1}^{L_z-z_1}\left(\frac{L_xx'^2}{L_y}
    +\frac{L_yy'^2}{L_x}
         +\frac{z^2}{L_xL_y}\right)^{-(d-2+\eta)/2},
   \eeq
where we have scaled the variables $x \to L_xx'$ and $y \to L_yy'$. In the thin-film structure,
we set $L_x = L_y = L$ and $L_z = l$, so that
   \beq
         k_BT_C\chi(T_C) \sim  \frac{(L^2)}{l}^{1-(d-2+\eta)/2}\int_0^1\mathrm{d}x'\int_0^1\mathrm{d}y'
         \sum_{z_1=1}^{l}\sum_{z=1-z_1}^{l-z_1}\left(x'^2+y'^2
         +\frac{z^2}{L^2}\right)^{-(d-2+\eta)/2}.
   \eeq
As $z \subset [1-l,l-1]$ and if we choose $l \ll L$, $\frac{z^2}{L^2} \ll 1$, this yields
$\sum_{z_1=1}^l\sum_{z=1-z_1}^{l-z_1}(x'^2+y'^2+\frac{z^2}{L^2}) \approx l^2(x'^2+y'^2)$. So
the susceptibility becomes
   \barr \label{eqn:chil_exponent}
         k_BT_C\chi(T_C) &\sim&
         l(L^2)^{1-\frac{d-2+\eta}{2}}\int_0^1\mathrm{d}x'\int_0^1\mathrm{d}y'(x'^2+y'^2)^{-(d-2+\eta)/2} \nonumber \\
    &\propto& L^{\gamma/\nu+2-d}l \quad \propto \quad L^{(\gamma/\nu)'},
   \earr
   where the scaling relation $\gamma/\nu = 2-\eta$ (e.g. \cite{Fisher1969})
has been used. Note that for a particular thin-film system, $l$ is a constant so the only
variable in the function is $L$. Consequently, in thin-films, if $l \ll L$, $\chi$ scales with
$L$  in the same way as isotropic-shaped systems, but with a different exponent $(\gamma/\nu)'
\equiv \gamma/\nu +2 -d $. For $<\!m\!>_{T_C}$, we may assume that it is of the same order as
the root mean square magnetisation $<\!m^2\!>^{1/2}_{T_C}$. So,
   \barr \label{eqn:ml_exponent}
      <\!m\!>_{T_C}  &\propto& <\!m^2\!>^{1/2}_{T_C}\  = \left[\frac{k_BT_C\chi(T_C)}{L^2l}\right]^{1/2}, \nonumber \\
      &\propto& L^{-\beta/\nu},
   \earr
where again we have used $\gamma/(2\nu)-d/2=-\beta/\nu$ \cite{Fisher1969}. Note that
since $\xi =\xi(l)$
in our hypothesis is a function of thickness $l$, every exponent calculated in
this way will also be a function of thickness.

\section{Mean-Field Approximation} \label{apd:mean_field}
In a mean-field study of thin-films, the average Weiss field is layer-dependent. We write down
the expression for the free energy for an $l$-layered films and obtain the equations of state
from its minimisation with respect to the layer-resolved magnetisations $m_k$ ($k$ is a layer
index). The $l$-coupled equations are solved numerically to extract the equilibrium
magnetisations $m_k$. The magnetisation per site  in the $k^{th}$ layer is given as $m_k =
P_{k,\uparrow}-P_{k,\downarrow}$ where $P_{k,(\uparrow,\downarrow)}$ is the probability of a
site being occupied by an up (down) spin. Including the interaction with an external field
($-h\sum_iS_i$) the interaction energy $U$ is the sum over all layers ($k= 1,\cdots,l$) of
terms $E_k$ which are written as
   \beq \label{eqn:E_i}
      E_k = -\frac{N_\parallel}{2}\left\{Z_0 J_{kk} m_k^2
     + Z_1J_{k,k+1}m_km_{k+1}(1-\delta_{k,l})+Z_1J_{k,k-1}m_km_{k-1}(1
     -\delta_{k,1})+2hm_k\right\},
   \eeq
where $Z_0$ and $Z_1$ are the number of nearest neighbors to a lattice site in the same layer
and one of its adjacent layers, and $N_\parallel$ is the number of sites in a layer. The free
boundary condition necessitates the presence of the factors $1-\delta_{k,1}$ and
$1-\delta_{k,l}$ in the equation. Similarly the entropy $S$ can be written as $S=
\sum_{k=1}^{l} S_k$ where
   \barr
      S_k &=& -k_B N_\parallel \left[P_{k,\uparrow}\ln P_{k,\uparrow}
      +(1-P_{k,\uparrow})\ln (1-P_{k\uparrow})\right]
      \nonumber \\
      &=& -k_BN_\parallel \left[\frac{1+m_k}{2}\ln\frac{1+m_k}{2}
      +\frac{1-m_k}{2}\ln\frac{1-m_k}{2}
      \right].
   \earr
Thus, by minimising the free energy $F=U-TS$ with respect to $m_k$ i.e. $\frac{\partial
F}{\partial m_k} = 0$, we obtain the equilibrium magnetisation in layer $k$ by solving the
following $l$-coupled equations for $l$-layered films
   \barr \label{eqn:meanfield_couple_equations}
      -(Z_0 J_{1,1} m_1 + Z_1 J_{1,2} m_2+h)
      + \frac{k_BT}{2}\ln \left[\frac{1+m_1}{1-m_1}\right] &=&0 ,\nonumber \\
      -(Z_0 J_{k,k} m_k + Z_1 J_{k,k+1} m_{k+1}+ Z_1 J_{k,k-1} m_{k-1}+h)
      + \frac{k_BT}{2}\ln \left[\frac{1+m_k}{1-m_k}\right]&=& 0,
      \quad k = 2,\cdots,l-1 \nonumber \\
      -(Z_0 J_{l,l} m_n + Z_1 J_{l, l-1} m_{l-1}+h)
      + \frac{k_BT}{2}\ln \left[\frac{1+m_l}{1-m_l}\right]
      &=& 0.
   \earr
$T_C$ can also be extracted from eqn. (\ref{eqn:meanfield_couple_equations}) for zero external
field. Close to $T_C$, the $\{m_i\}$ are small and $\ln\frac{1+m_k}{1-m_k}\approx 2m_i$ and
$T_C$ can be extracted by finding the temperature for which the set of equations
$\mathbf{A}\mathbf{m} = \mathbf{0}$, where $\mathbf{A}$ is a $l\times l$ matrix with elements
   \beq
      A_{ij} = (k_BT - Z_0 J_{i,j})\delta_{i,j}-Z_1J_{i,j}\left\{(1-\delta_{i,1})
      \delta_{i,j-1}+(1-\delta_{i,l})\delta_{i,j+1}\right\},
   \eeq
and $\mathbf{m}$ is a $l\times 1$ column matrix,$\{m_1,\cdots,m_l\}$, is satisfied.


\begin{table}[h*]
   \begin{small}
   \begin{tabular}{|c|c|cc|c|c|c||} \hline\hline
         &\textbf{\#layers}   & \multicolumn{2}{c|}{$T_C$} & $\beta/\nu$ & $\gamma/\nu+2-d$ & $1/\nu$ \\
         &                    &  MC    &  MF               &             &                  & \\\hline\hline
         &2D(Exact)   & $\approx$2.269185 &\ 4 & 0.125 & 1.75 & 1  \\\hline\hline
         &1           &  $2.2693 \pm 3\times 10^{-4}$ &\ 4       & $0.126 \pm 1\times 10^{-3}$ & $1.753 \pm 8\times 10^{-3}$ & $1.01 \pm 2\times 10^{-2}$\\
         &2           &  $3.2076 \pm 4\times 10^{-4}$ &\ 5       & $0.126 \pm 7\times 10^{-3}$ & $1.750 \pm 7\times 10^{-3}$ & $1.00 \pm 1\times 10^{-2}$\\
         &4           &  $3.8701 \pm 3\times 10^{-4}$ &\ 5.61803 & $0.125 \pm 2\times 10^{-3}$ & $1.730 \pm 1\times 10^{-2}$ & $1.00 \pm 2\times 10^{-2}$\\
         &6           &  $4.1179 \pm 3\times 10^{-4}$ &\ 5.80194 & $0.131 \pm 2\times 10^{-3}$ & $1.740 \pm 1\times 10^{-2}$ & $1.02 \pm 2\times 10^{-2}$\\
      sc &8           &  $4.2409 \pm 2\times 10^{-4}$ &\ 5.87939 & $0.141 \pm 2\times 10^{-3}$ & $1.744 \pm 7\times 10^{-3}$ & $1.03 \pm 1\times 10^{-2}$\\
         &10          &  $4.3117 \pm 3\times 10^{-4}$ &\ 5.91899 & $0.144 \pm 2\times 10^{-3}$ & $1.720 \pm 1\times 10^{-2}$ & $1.00 \pm 3\times 10^{-2}$\\
         &15          &  $4.3996 \pm 4\times 10^{-4}$ &\ 5.96157 & $0.162 \pm 2\times 10^{-3}$ & $1.680 \pm 1\times 10^{-2}$ & $1.02 \pm 1\times 10^{-2}$\\
         &20          &  $4.4381 \pm 2\times 10^{-4}$ &\ 5.97766 & $0.166 \pm 3\times 10^{-3}$ & $1.660 \pm 2\times 10^{-2}$ & $1.03 \pm 1\times 10^{-2}$\\
         &Bulk        &  $4.5115 \pm 2\times 10^{-4}$ &\ 6 & $0.507 \pm 4\times 10^{-3}$ & $0.980 \pm 1\times 10^{-2}$ & $1.57 \pm 3\times 10^{-2}$\\
         &Bulk$^*$    &  $4.5390 \pm 6\times 10^{-3}$ & $6.00200\pm 2\times 10^{-4}$ &-&-&-\\
         \hline\hline
         &2           &  $2.2691 \pm 1\times 10^{-4}$ &\ 4       & $0.120 \pm 5\times 10^{-3}$ & $1.760 \pm 3\times 10^{-2}$ & $1.01 \pm 1\times 10^{-2}$\\
         &4           &  $4.2947 \pm 3\times 10^{-4}$ &\ 6.47214 & $0.124 \pm 5\times 10^{-3}$ & $1.760 \pm 4\times 10^{-2}$ & $1.00 \pm 1\times 10^{-2}$\\
         &6           &  $5.1023 \pm 4\times 10^{-4}$ &\ 7.20775 & $0.130 \pm 6\times 10^{-3}$ & $1.762 \pm 6\times 10^{-3}$ & $1.02 \pm 1\times 10^{-2}$\\
         &8           &  $5.5005 \pm 3\times 10^{-4}$ &\ 7.51754 & $0.131 \pm 1\times 10^{-3}$ & $1.752 \pm 7\times 10^{-3}$ & $1.02 \pm 1\times 10^{-2}$\\
      bcc&10          &  $5.7287 \pm 2\times 10^{-4}$ &\ 7.67594 & $0.136 \pm 1\times 10^{-3}$ & $1.748 \pm 7\times 10^{-3}$ & $1.04 \pm 1\times 10^{-2}$\\
         &15          &  $6.0076 \pm 4\times 10^{-4}$ &\ 7.84628 & $0.153 \pm 2\times 10^{-3}$ & $1.730 \pm 1\times 10^{-2}$ & $1.06 \pm 1\times 10^{-2}$\\
         &20          &  $6.1300 \pm 3\times 10^{-4}$ &\ 7.91065 & $0.169 \pm 3\times 10^{-3}$ & $1.700 \pm 1\times 10^{-2}$ & $1.06 \pm 1\times 10^{-2}$\\
         &Bulk        &  $6.3557 \pm 1\times 10^{-4}$ &\ 8       & $0.505 \pm 3\times 10^{-3}$ & $1.001 \pm 4\times 10^{-3}$ & $1.60 \pm 1\times 10^{-2}$\\
         &Bulk$^*$    &  $6.3890 \pm 3\times 10^{-3}$ & $8.03200\pm 4\times 10^{-3}$ &-&-&-\\
         \hline\hline
         &1           &  $2.2690 \pm 2\times 10^{-4}$ &\ 4       & $0.124 \pm 3\times 10^{-3}$ & $1.740 \pm 2\times 10^{-2}$ & $1.00 \pm 2\times 10^{-2}$\\
         &2           &  $5.2416 \pm 6\times 10^{-4}$ &\ 8       & $0.126 \pm 3\times 10^{-3}$ & $1.751 \pm 6\times 10^{-3}$ & $0.99 \pm 1\times 10^{-2}$\\
         &4           &  $7.5702 \pm 3\times 10^{-4}$ & 10.47210 & $0.126 \pm 1\times 10^{-3}$ & $1.750 \pm 1\times 10^{-2}$ & $0.99 \pm 3\times 10^{-2}$\\
         &6           &  $8.4431 \pm 3\times 10^{-4}$ & 11.20780 & $0.128 \pm 1\times 10^{-3}$ & $1.751 \pm 4\times 10^{-3}$ & $1.01 \pm 1\times 10^{-2}$\\
      fcc&8           &  $8.8679 \pm 7\times 10^{-4}$ & 11.51750 & $0.130 \pm 2\times 10^{-3}$ & $1.740 \pm 1\times 10^{-2}$ & $1.00 \pm 1\times 10^{-2}$\\
         &10          &  $9.1106 \pm 6\times 10^{-4}$ & 11.67590 & $0.138 \pm 3\times 10^{-3}$ & $1.730 \pm 1\times 10^{-2}$ & $1.00 \pm 2\times 10^{-2}$\\
         &15          &  $9.4057 \pm 3\times 10^{-4}$ & 11.84630 & $0.153 \pm 2\times 10^{-3}$ & $1.720 \pm 1\times 10^{-2}$ & $1.03 \pm 2\times 10^{-2}$\\
         &20          &  $9.5353 \pm 5\times 10^{-4}$ & 11.91060 & $0.173 \pm 3\times 10^{-3}$ & $1.710 \pm 1\times 10^{-2}$ & $1.07 \pm 2\times 10^{-2}$\\
         &Bulk        &  $9.7736 \pm 2\times 10^{-4}$ & 12       & $0.498 \pm 8\times 10^{-3}$ & $0.994 \pm 9\times 10^{-3}$ & $1.59 \pm 3\times 10^{-2}$\\
         &Bulk$^*$    &  $9.8600 \pm 2\times 10^{-2}$ & $12.04500\pm 6\times 10^{-3}$ &-&-&-\\
         \hline\hline
   \end{tabular}
   \end{small}
      \caption{ The critical temperatures determined by both Monte Carlo (MC) simulations and also
      mean-field (MF) calculations for Ising films. Bulk $3d$ results are also given for
      comparison. Bulk$^{\ast}$ refers to the $T_C (\infty)$ parameter of the fitting expression
      of eqn. (\ref{eqn:fitting_bulk_Tc}). Critical exponents are given which have been extracted
      from the MC simulations via a finite size scaling analysis. Since we can extract the ratio
      of exponents $\gamma/\nu$ for the bulk $3d$ system we give $(\gamma/\nu)'$ as
      $\gamma/\nu +2 -3$}
      \label{table:mc_ising_thin_films}
\end{table}

\begin{table}[h*]
   \begin{small}
   \begin{tabular}{||c|c|c|c|c|c|c||} \hline\hline
         & Structure  & $T_C(\infty)$    &     $l_0$    &     $l'$     &     $\lambda'$   &  $\lambda(\infty) = 1/\nu^{3d}$ \\
         \hline
         & sc  & $4.539\pm 6\times 10^{-3}$ & $1.010\pm 2\times 10^{-2}$ & $-0.010\pm 2\times 10^{-2}$ & $1.280\pm 2\times 10^{-2}$ & $1.578\pm 3\times 10^{-3}$ \\
      MC & bcc & $6.389\pm 3\times 10^{-3}$ & $1.933\pm 7\times 10^{-3}$ & $ 0.746\pm 7\times 10^{-3}$ & $1.380\pm 6\times 10^{-3}$ & $1.620\pm 1\times 10^{-2}$ \\
         & fcc & $9.860\pm 2\times 10^{-2}$ & $1.416\pm 6\times 10^{-3}$ & $ 0.440\pm 7\times 10^{-3}$ & $1.300\pm 1\times 10^{-2}$ & $1.621\pm 8\times 10^{-3}$ \\
         \hline
         & sc  & $6.0020\pm 2\times 10^{-4}$ & $1.117\pm 3\times 10^{-3}$ & $-0.61\pm 4\times 10^{-2}$ & $1.894\pm 3\times 10^{-3}$ & $2.002\pm 4\times 10^{-3}$ \\
      MF & bcc & $8.0320\pm 4\times 10^{-3}$ & $1.380\pm 1\times 10^{-3}$ & $ 0.63\pm 2\times 10^{-2}$ & $1.600\pm 6\times 10^{-2}$ & $2.002\pm 2\times 10^{-3}$ \\
         & fcc &$12.0500\pm 6\times 10^{-3}$ & $1.120\pm 1\times 10^{-2}$ & $ 0.28\pm 1\times 10^{-2}$ & $1.580\pm 1\times 10^{-2}$ & $2.000\pm 2\times 10^{-3}$ \\
         \hline\hline
   \end{tabular}
   \end{small}
      \caption{Fitting parameters for Ising thin-films  using eqn.
      (\ref{eqn:fitting_bulk_Tc}). MC and MF stand for Monte Carlo and mean-field respectively.}
      \label{table:mc_mf_fitting_parameters}
\end{table}

   \bef[h*]
      \includegraphics{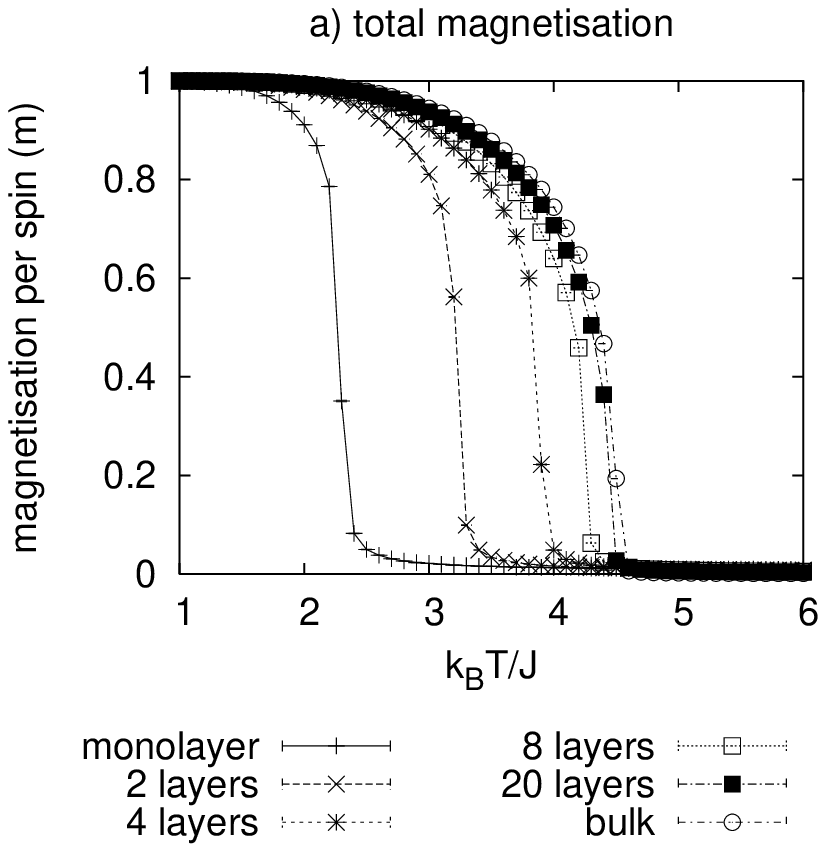}
      \includegraphics{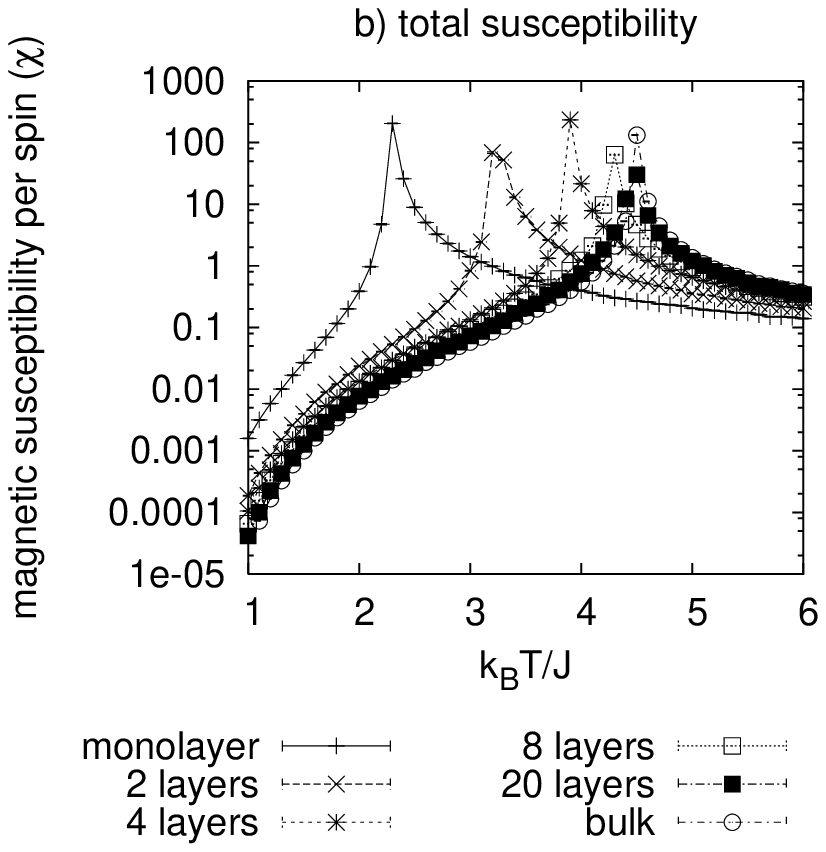}
      \includegraphics{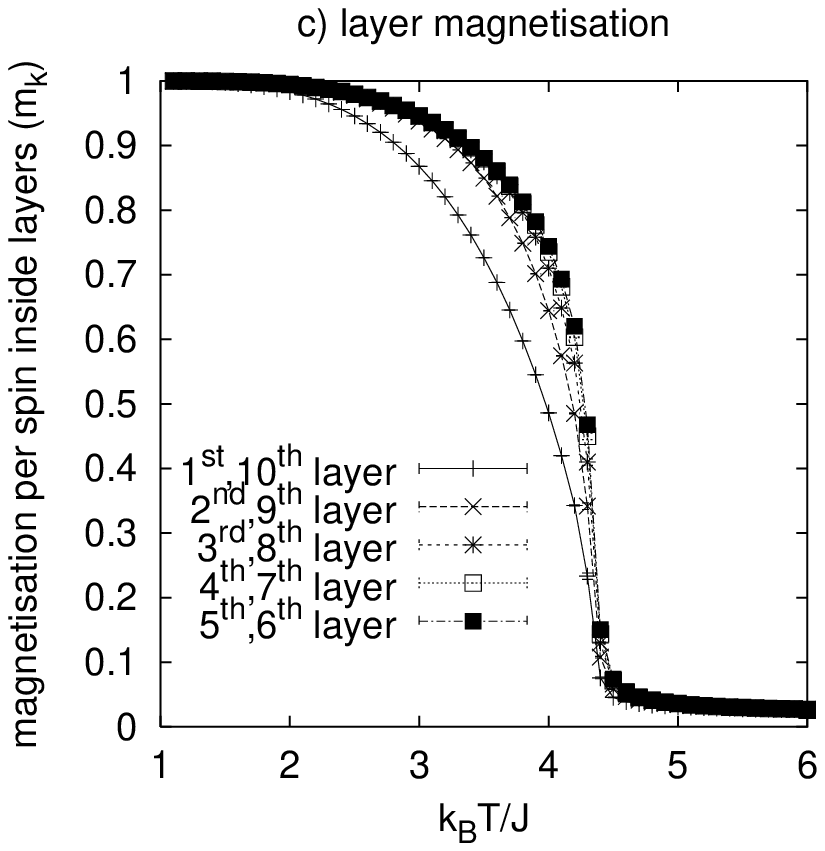}
      \includegraphics{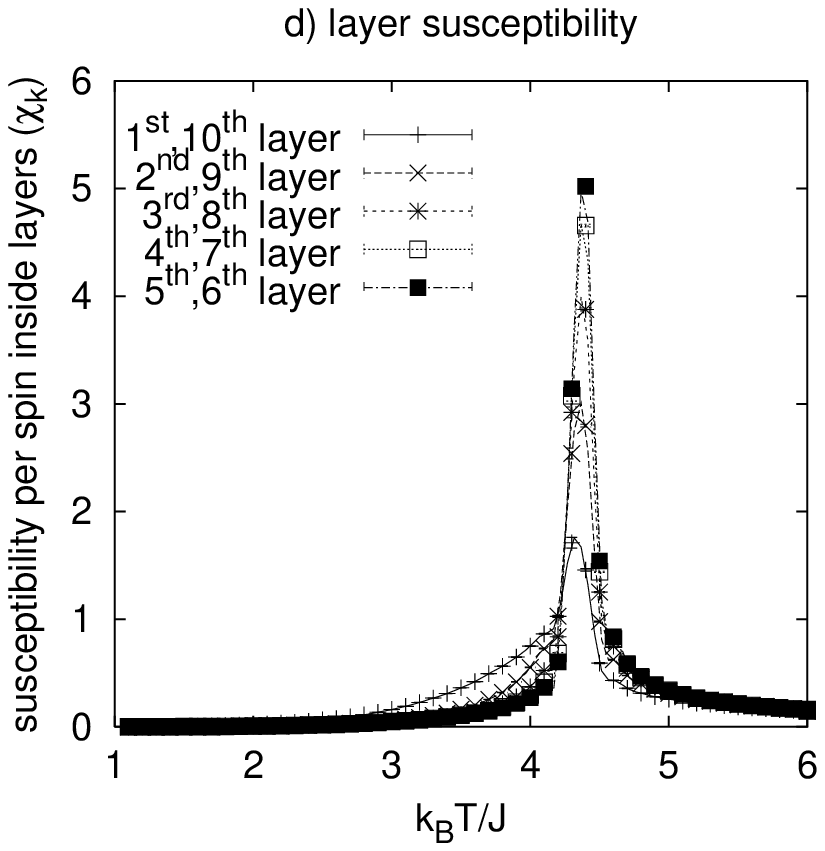}
      \caption{The magnetic properties of Ising sc thin-films from Monte Carlo (MC)
      simulations. (a) and (b) show a dimensional crossover of  $m$ and $\chi$ respectively
      from $2d$- to bulk-like for $128 \times 128 \times l$ spins ($l$ is
      the number of layers in the film and $l = 128$ represents the bulk
      system). (c) and (d) show the layer-dependence of $m_k$ and $\chi_k$,
      where $k$ is a layer index,
      in $128\times 128\times 10$ sc films. Owing to the
      isotropic exchange interaction and spatial symmetry, layers $k$ and
      $10-k$ have the same properties. Lines are added as a viewing aid.
      The error bars are smaller than point size.}
      \label{figure:mc_ising_sc_films_magnetic_properties_crossover}
   \eef

   \bef[h*]
      \includegraphics{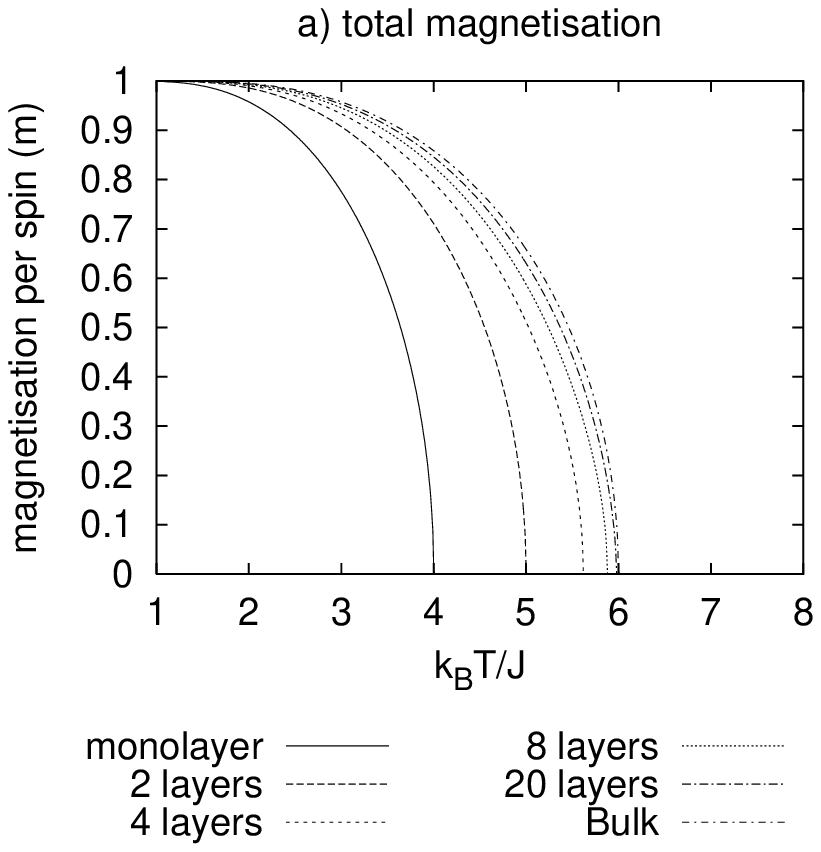}
      \includegraphics{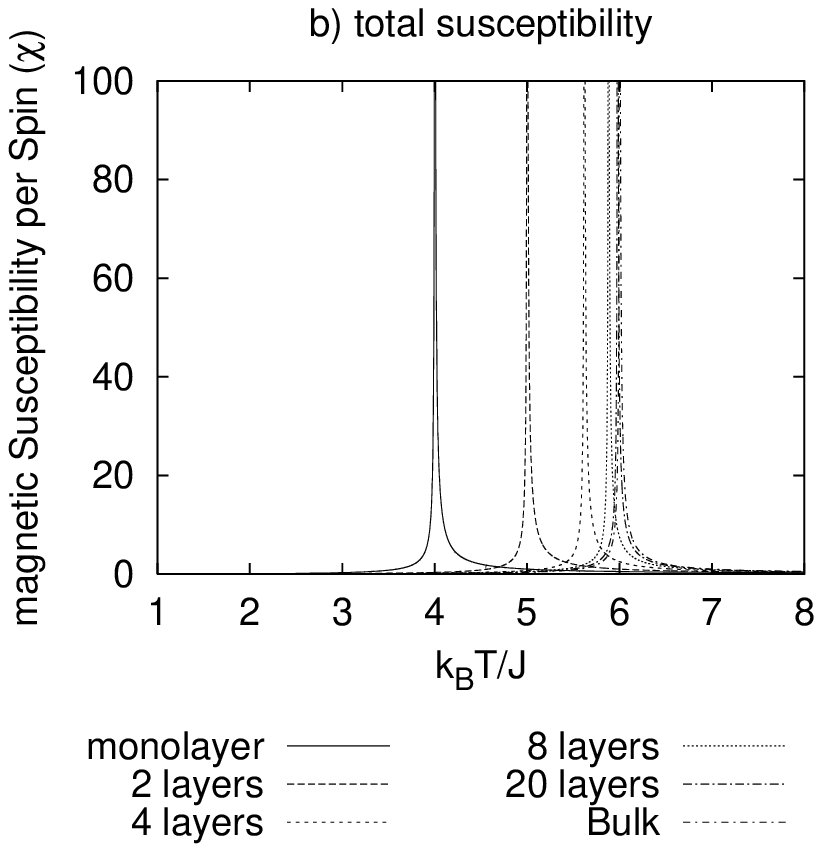}
      \includegraphics{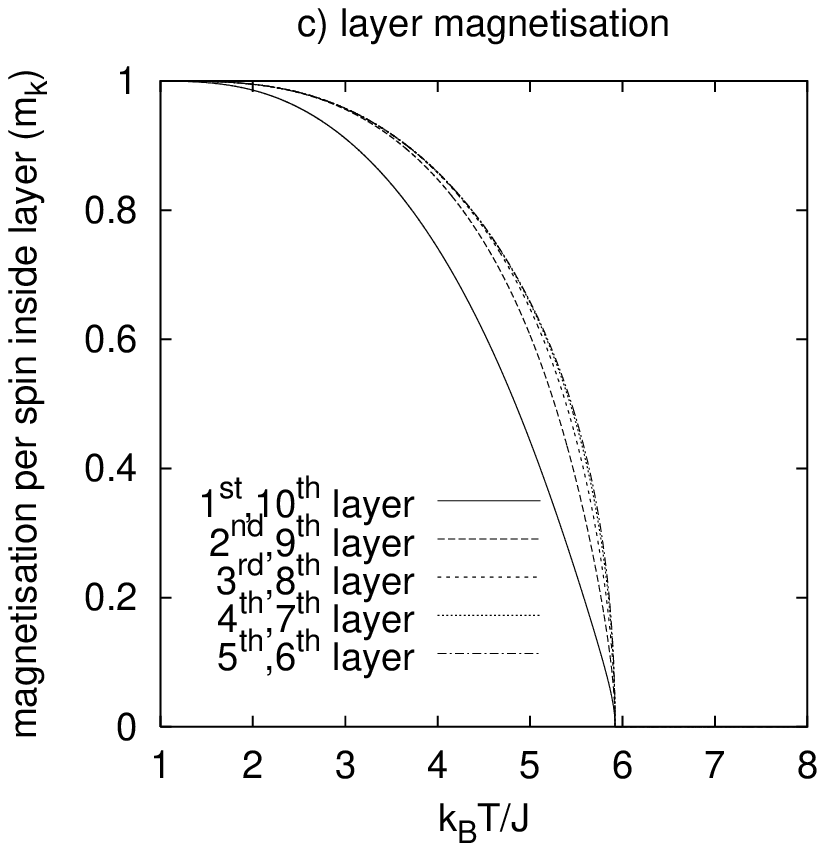}
      \includegraphics{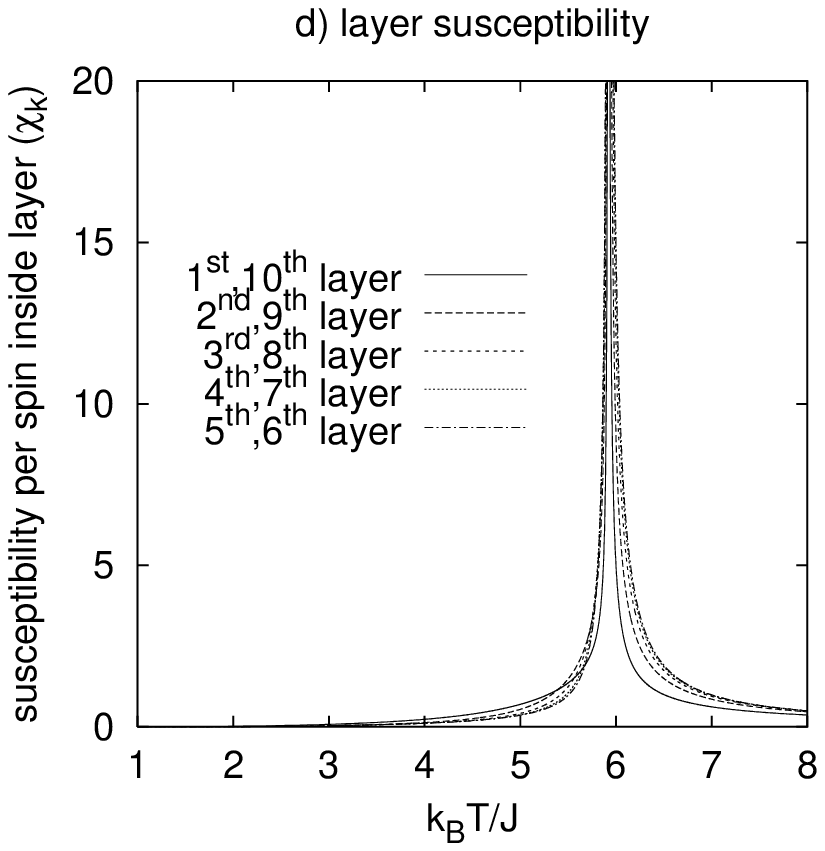}
      \caption{The magnetic properties of Ising sc thin-films from mean-field calculations.
      As in Fig. \ref{figure:mc_ising_sc_films_magnetic_properties_crossover},
      (a) and (b) show a crossover of $m$ and $\chi$ from the $2d$ to the bulk limit
      ($k_BT_C/J$ varies from 4 to 6).
      (c) and (d) show examples of the layer dependence of $m_k$ and $\chi_k$ for
      10-layered sc films.}
      \label{figure:mf_ising_sc_films_magnetic_properties_crossover}
   \eef

   \bef[h*]
       \includegraphics{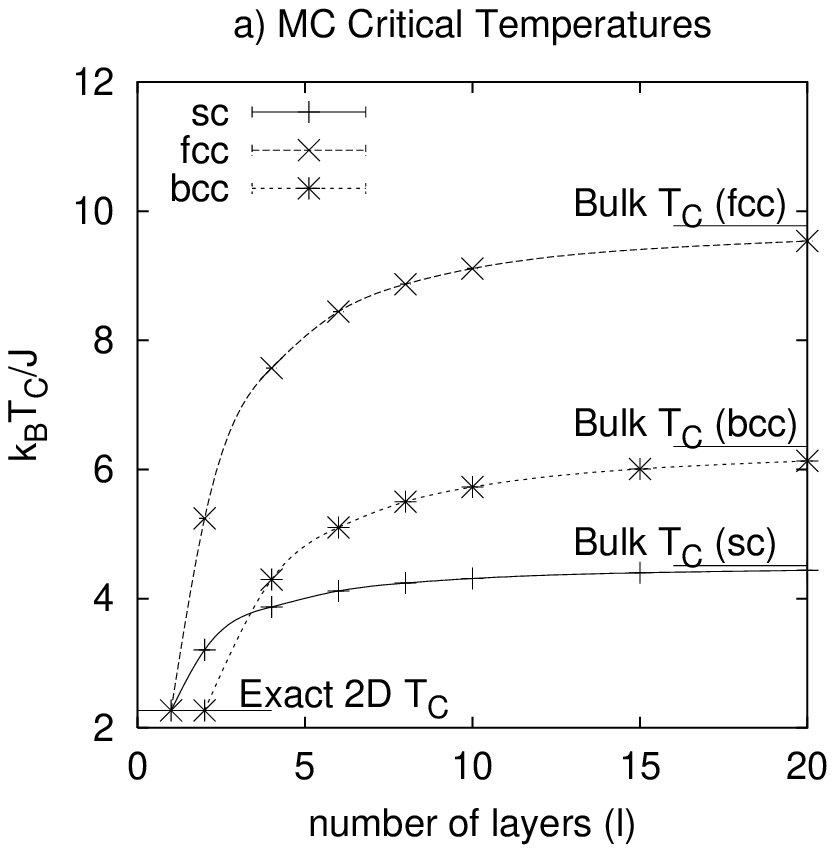}
       \includegraphics{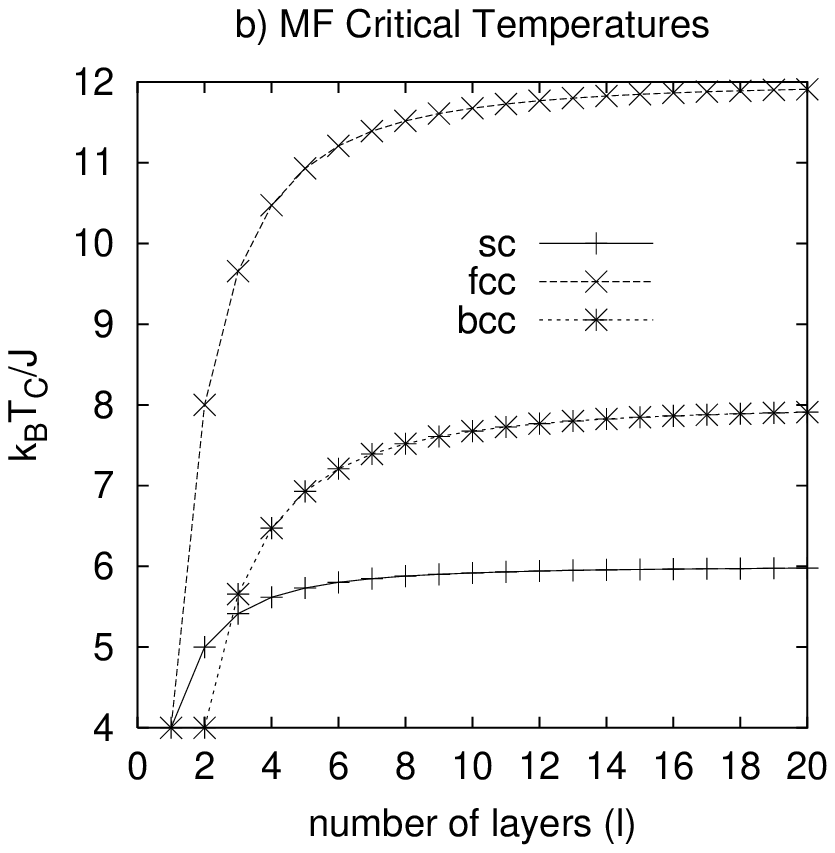}
       \caption{The critical temperatures $T_C$ as a function of thickness $l$
       extracted from (a) Monte Carlo simulations and (b) mean-field calculations.
       For mean-field, the values of $k_BT_C/J$ are 4 for $2d$ and
       6, 8 and 12 for bulk sc, bcc, and fcc respectively. Lines are added
       as a viewing aid.  The error bars in (a) are smaller than point size.}
       \label{figure:mc_mf_ising_Tc}
   \eef

   \bef[h*]
      \begin{minipage}[h*]{8.5cm}
         \includegraphics{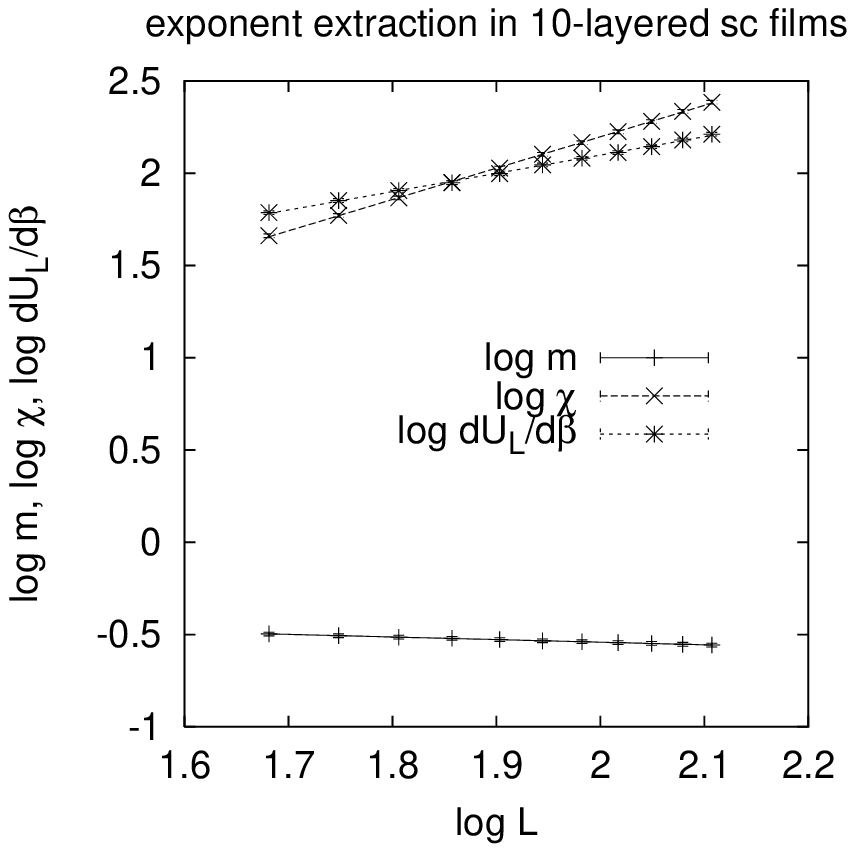}
         \caption{An example of the extraction of critical exponents $\beta$,$(\gamma/\nu)'$, and $1/\nu$ from the
         slope of a least-square-fit (see text) for 10-layered sc films. The apparent
         linear relation supports  eqn. (\ref{eqn:mchi_scaling_in_thin_films})}
         \label{figure:mc_ising_sc_films_exponent_extraction}
      \end{minipage}
      \hspace{0.5cm}
      \begin{minipage}[h*]{8.5cm}
         \includegraphics{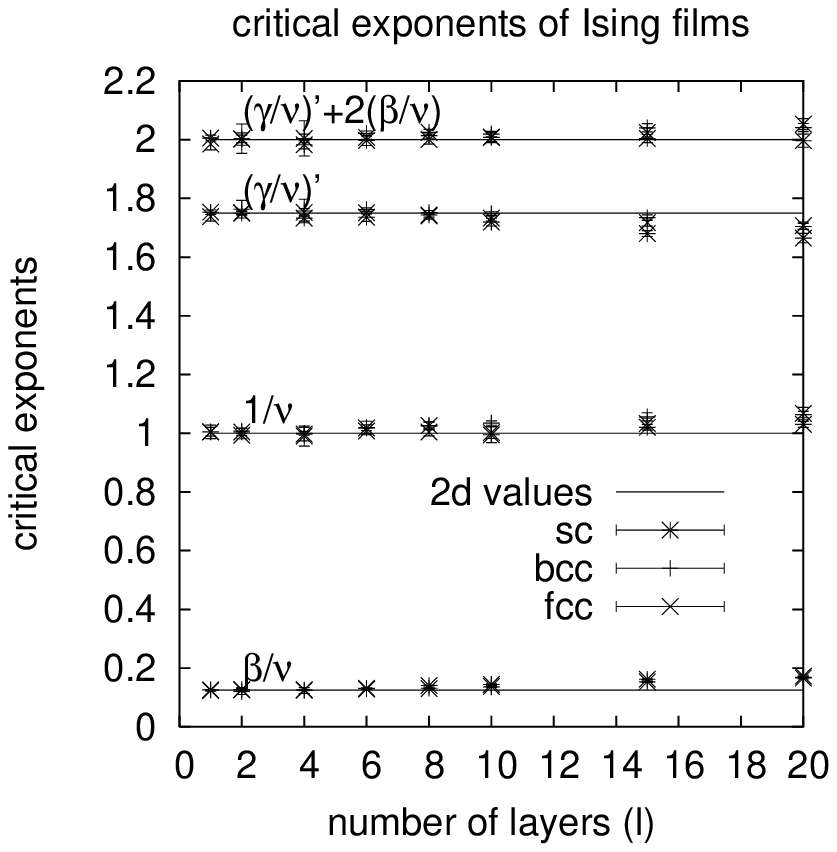}
         \caption{Results of critical exponents for all sc, bcc, and fcc films. In all structures
         the exponents are very close to $2d$ values. Especially, for thin-films ($l \leq 8$
         approximately), the exponents are consistent with $2d$ results and suggest that thin-films belong
         to the $2d$ class. The summation $(\gamma/\nu)'+2\beta/\nu) = 2$ supports our single
         correlation length $\xi$ assumption.}
         \label{figure:mc_ising_sc_films_exponent}
       \end{minipage}
   \eef

   \bef[h*]
      \includegraphics{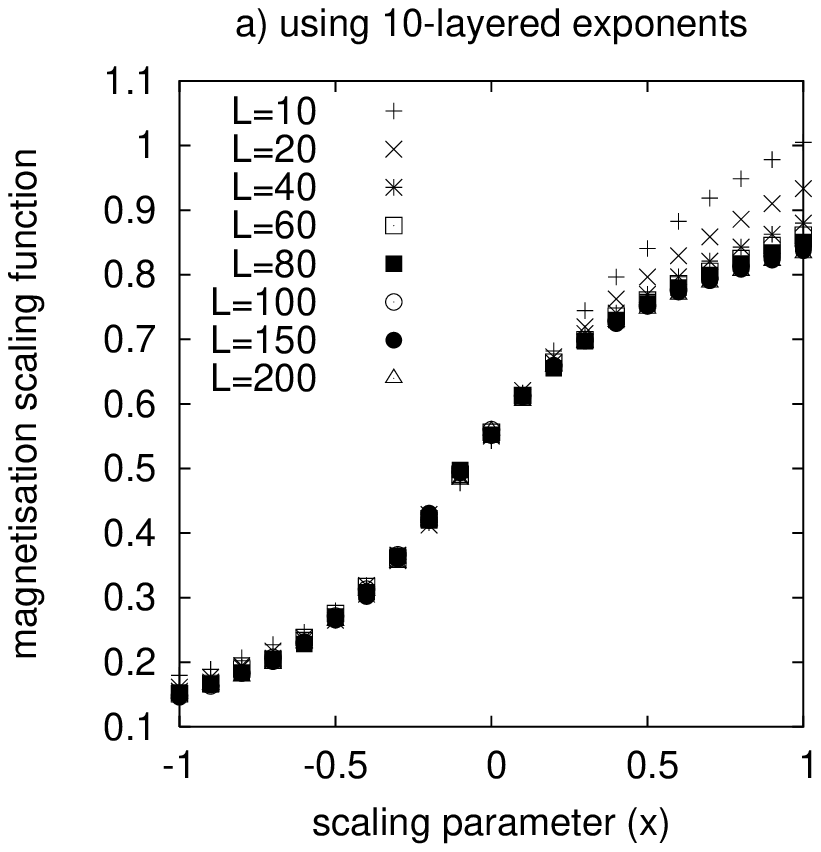}
      \includegraphics{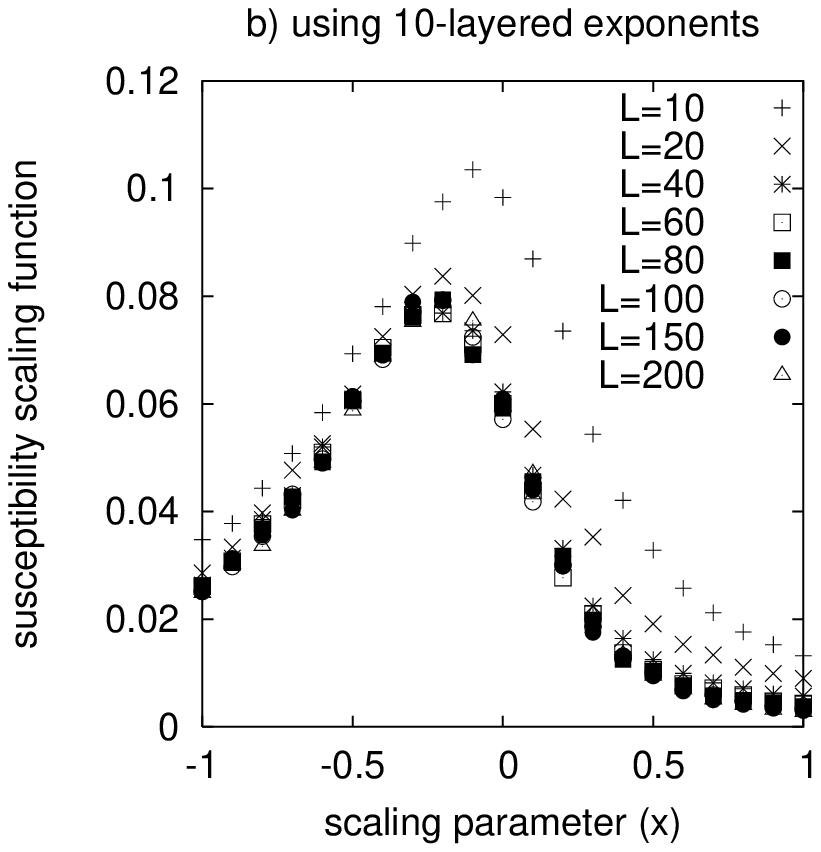}
      \includegraphics{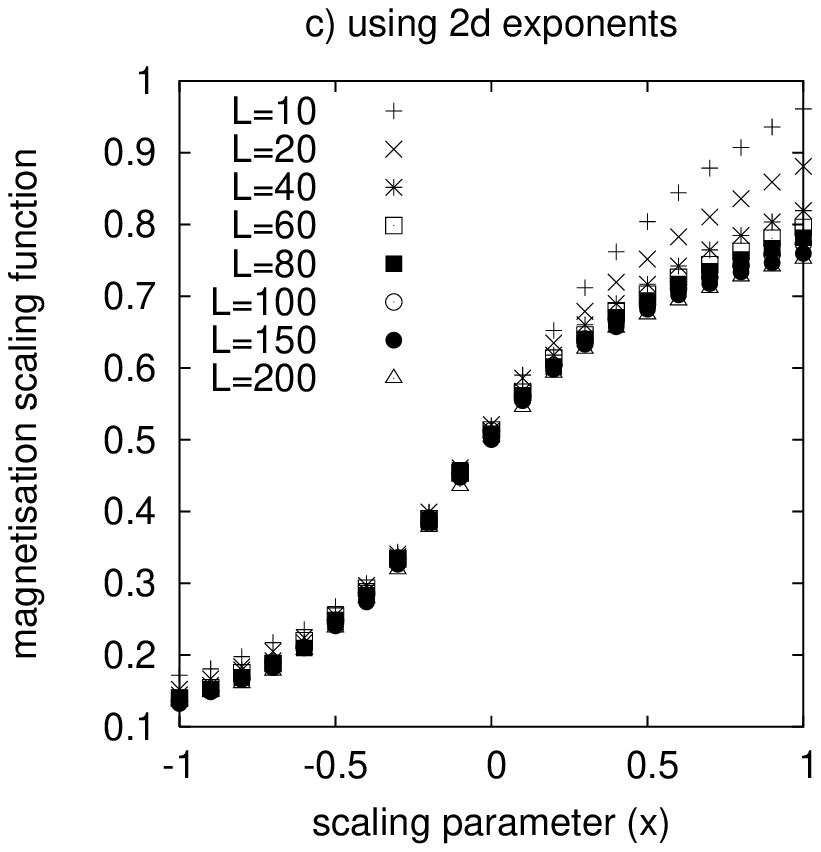}
      \includegraphics{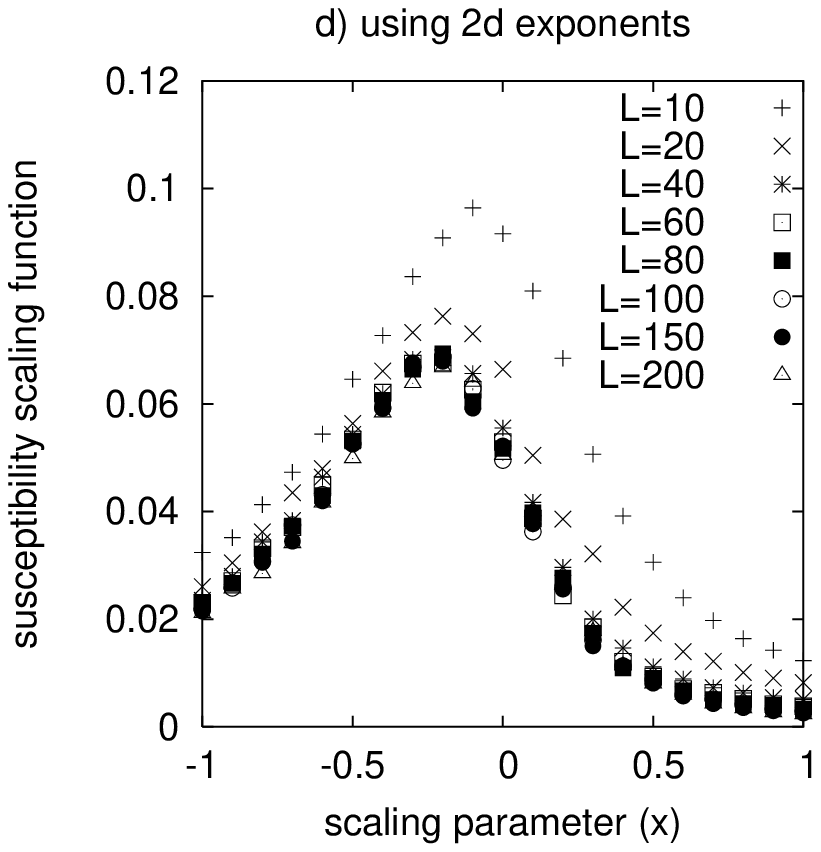}
      \caption{The scaling functions for the magnetisation (a) and susceptibility (b)
      in 10 layered sc systems with $L$ from 100 to 200
      calculated by using those critical exponents extracted for 10-layered
      sc films with $L$ ranging from $64$ to $128$. Analogous plots using
      $2d$ critical exponents are shown in (c) and (d). In (a) and (b), we find a
      good collapse of the data for all systems with $L \geq 60$ including
      $L = 150, 200$ which are bigger than those used in the exponent extraction. Note, however, that
      for $L$ close to the thickness $l = 10$, the collapsing is not found
      since the  assumption underpinning our finite size scaling relation
      (eqn. \ref{eqn:mchi_scaling_in_thin_films}) rests on the
      condition $L \gg l$. In (c) and (d) we find a slightly worse collapsing than that
      in (a) and (b) - compare, for example, the collapsing around $x=1$ in (c) and (a). This
      confirms the weak $l$ dependence of critical exponents and the deviation from the $2d$
      values when the films thicken.}
      \label{figure:mf_ising_sc_scaling_functions}
   \eef

   \bef[h*]
      \begin{minipage}[h*]{8.5cm}
         \includegraphics{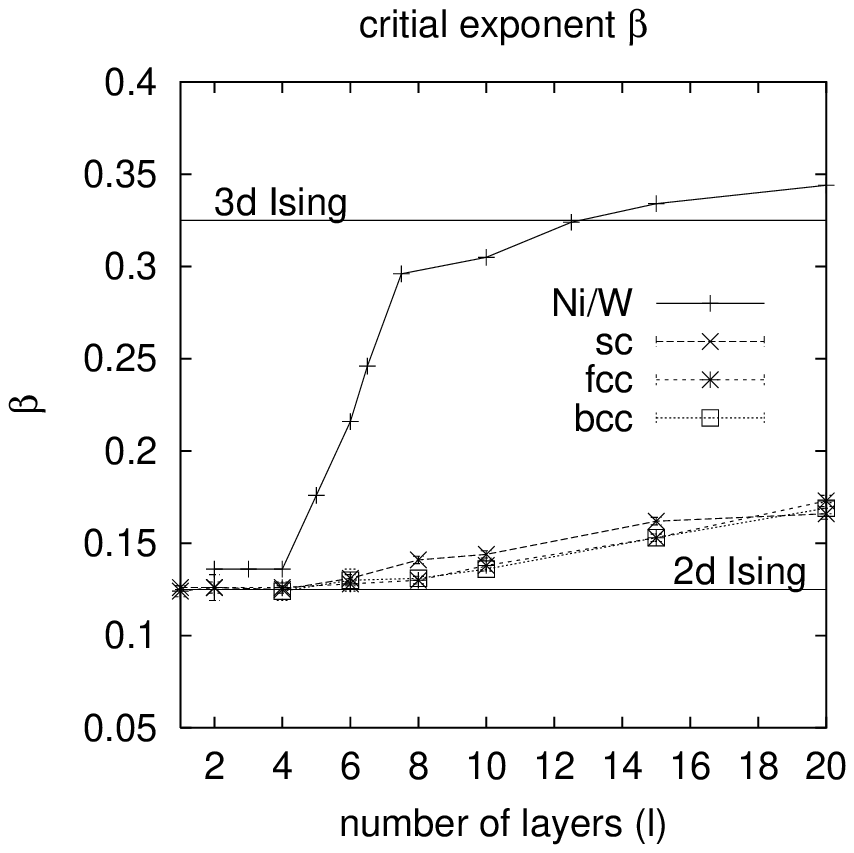}
         \caption{Comparison of the critical exponent $\beta$ extracted from
         experimental data for Ni(111)/W(110) in \cite{Li1992} and from Monte Carlo
         simulations of Ising thin-films.}
         \label{figure:mc_ising_beta_comparison}
      \end{minipage}
      \hspace{0.5cm}
      \begin{minipage}[h*]{8.5cm}
         \includegraphics{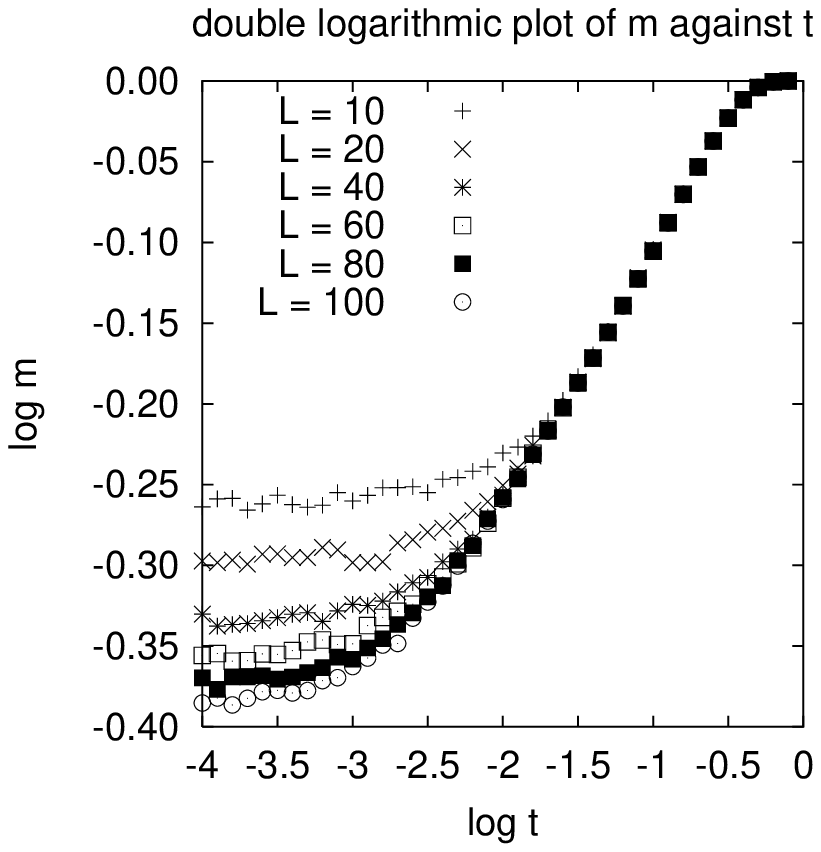}
         \caption{Double logarithmic plot of $m$ against $t$, the reduced
    temperature, for 4-layered Ising sc films. The finite size effect is
         evident once $\xi$ becomes comparable to the size $L$ of the layers
    of the films. Note the data is reliable down to
    $\log t \approx -\frac{1}{\nu}\log L$ }
         \label{figure:mc_ising_logm_vs_logt}
      \end{minipage}
   \eef

   \bef[h*]
      \begin{center}
      \includegraphics{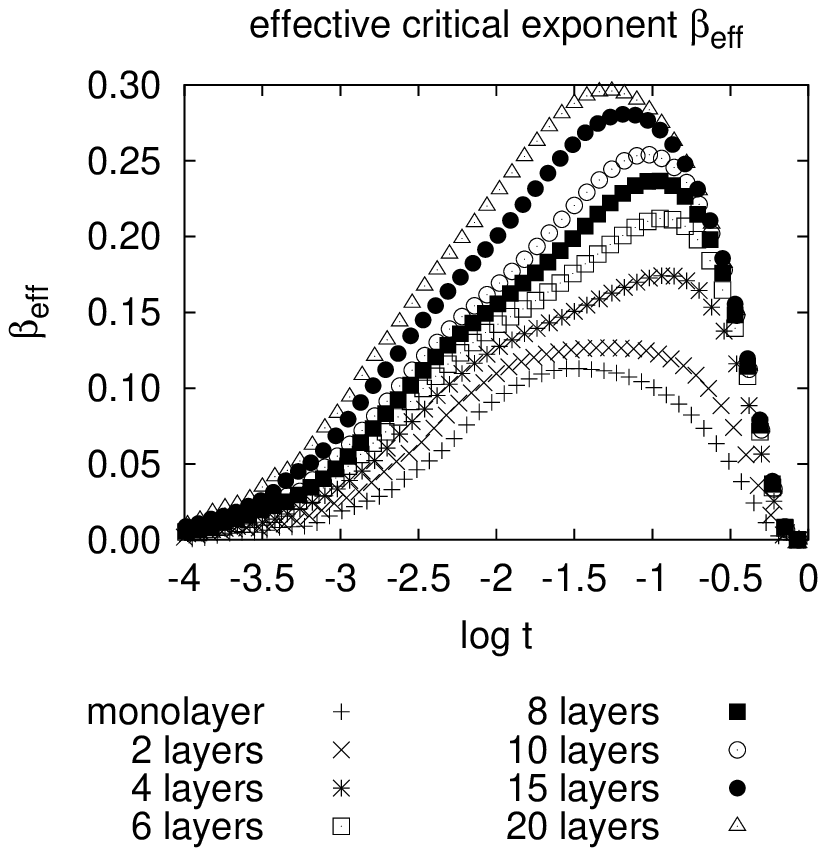}
      \end{center}
      \caption{The effective critical exponent $\beta_{\mathrm{eff}}$ extracted from
      $100\times 100\times l$ sc Ising films
      where $l$ is the number of layers. For $l \geq 4$, it is found that around $\log t = -1$
      to $-1.5$, the $\beta_{\mathrm{eff}}$ curves reach peaks and this indicates how a
      step-like function of $\beta$ of the form described in \cite{Li1992}
      i.e. Fig. \ref{figure:mc_ising_beta_comparison} is possible.}
      \label{figure:mc_ising_sc_beta_eff}
   \eef

\begin{thebibliography}{99}
   \bibitem{Falicov1990} L.M. Falicov, D.T. Pierce, S.D. Bader, R. Gronsky, K.B. Hathaway, H.J.
      Hopster, D.N. Lambeth, S.S.P. Parkin, G. Prinz, M. Salamon, I.K. Schuller, and R.H. Victora,
      J. Mater. Res. {\bf 5}, 1299 (1990).
   \bibitem{Johnson1996} M.T. Johnson, P.J.H. Bloemen, F.J.A. den Broeder, and J.J. de Vries,
      Rep. Prog. Phys. {\bf 59}, 1409 (1996).
   \bibitem{Li1992} Y. Li and K. Baberschke, Phys. Rev. Lett. {\bf 68}, 1208 (1992).
   \bibitem{Binder1994} K. Binder and P.C. Hohenberg, Phys. Rev. B {\bf 9}, 2194 (1974).
   \bibitem{Bander1988} M. Bander and D.L. Mills, Phys. Rev. B {\bf 38}, 12015 (1998).
   \bibitem{Elmers1994} H.J. Elmers, J. Hauschild, H. H\"{o}che, U. Gradmann, H. Bethge, D.
      Heuer and U. K\"{o}hler, Phy. Rev. Lett. {\bf 73}, 898 (1994).
   \bibitem{Allen1970} G.A.T. Allan, Phys. Rev. B {\bf 1}, 352 (1970).
   \bibitem{Capehart1976} T.W. Capehart and M.E. Fisher, Phys. Rev. B {\bf 13}, 5021 (1976).
   \bibitem{O'Connor1994} D. O'Connor and C.R. Stephens, Phys. Rev. Lett. {\bf 72}, 506 (1994).
   \bibitem{Lin1992} D.L. Lin, H. Che and Y. Xia, Phys. Rev. A {\bf 46}, 1805 (1992); \\
      D.L. Lin, H. Che, W. Lai and T.F. George, Phys. Rev. E {\bf 49}, 2155 (1994); \\
      Y. Song, Y. Chen, J. Luo and D. Xian, Phys. Lett. A {\bf 221}, 124 (1996); \\
      J.T. Ou, F. Wang and D.L. Lin, Phys. Rev. E {\bf 56}, 2805 (1997).
   \bibitem{Binder1974} K. Binder, Thin Solid Films {\bf 20}, 367 (1974).
   \bibitem{Schilbe1996} P. Schilbe, S. Siebentritt, K.H. Rieder, Phys. Lett. A {\bf 216}, 20 (1996);\\
      P. Schilbe and K.H. Rieder, Europhys. Lett. {\bf 41}, 219 (1998); \\
      M.I. Marqu\'{e}s and J.A. Gonzalo, Eur. Phys. J. B {\bf 14}, 317 (2000); \\
      M.I. Marqu\'{e}s and J.A. Gonzalo, Nanotechnology {\bf 12}, 143 (2001).
   \bibitem{Wolff1989} U. Wolff, Phys. Rev. Lett. {\bf 62}, 361 (1989).
   \bibitem{Muller1973} H. M\"{u}ller-Krumbhaar and K. Binder, J. Stat. Phys. {\bf 8}, 1 (1973).
   \bibitem{Hohenberg1977} P.C. Hohenberg and B.I. Halperin, Rev. Mod. Phys. {\bf 49}, 435  (1977).
   \bibitem{L'Ecuyer1998} P. L'Ecuyer, in: {\it Handbook of Simulation}, ed. J. Banks (Wiley, 1998).
   \bibitem{Entacher1997} K. Entacher. {\it A collection of selected pseudorandom number generators with linear structures.}
      Technical report, Dept. of Mathematics University of Salzburg (1997).
   \bibitem{Coddington1994} P.D. Coddington, Int. J. Mod. Phys. C {\bf 5}, 547 (1994).
   \bibitem{Binder1981} K. Binder, Z. Phys. B {\bf 43}, 119 (1981).
   \bibitem{Ferrenberg1988} A.M. Ferrenberg and R.H. Swendsen, Phys. Rev. Lett. {\bf 61}, 2635 (1988).
   \bibitem{Ferrenberg1991} A.M. Ferrenberg and D.P. Landau, Phys. Rev. B {\bf 44}, 5081 (1991).
   \bibitem{Newman1999} M.E.J. Newman and G.T. Barkema, {\it Monte Carlo Methods in Statistical
      Physics} (Oxford University Press, Oxford, 1999);
   \bibitem{Privman1990} V. Privman, in: {\it Finite size scaling and Numerical Simulation of
      Statistical Systems}, ed. V. Privman (World Scientific, Singapore,
      1990 reprinted 1998).
   \bibitem{Binder1990} K. Binder, in: {\it Finite size scaling and numbercal simulation of
      statistical systems}, ed. V. Privman (World Scientific, Singapore, 1990 reprinted 1998).
   \bibitem{Stanley1971} H.E. Stanley, {\it Introduction to Phase Transitions and Critical
      Phenomena} (Clarendon Press, Oxford , 1971).
   \bibitem{Fisher1971} M.E. Fisher, in: {\it Critical Phenomena, Proc. Int. School of Physics,
      ``Enrico Fermi"}, Varenna, Italy, Course LI, ed. M.S. Green (Academic Press, New York,
      1971).
   \bibitem{Barber1983} M.N. Barber, in: {\it Phase Transitions and Critical Phenomena} Vol. 8,
      ed. C. Domb and J.L. Lebowitz (Academic Press Limited, London, 1983).
   \bibitem{Fisher1974} M.E. Fisher, Rev. Mod. Phys. {\bf 46}, 597 (1974).
   \bibitem{Fisher1969} M.E. Fisher, Phys. Rev. {\bf 180}, 594 (1969).
   \bibitem{Adler1983} J. Adler, J. Phys. A {\bf 16}, 3585 (1983) {\it and references contained therein}.
   \bibitem{Haubenreisser1972} W. Haubenreisser, W. Brodkorb, A. Corciovei and G. Costache, Phys.
   Status Solidi b {\bf 53}, 9 (1972) {\it and references contained therein}.
   \bibitem{Huang1994} F. Huang, M.T. Kief, G.J. Mankey, and R.F. Willis, Phys. Rev. B {\bf 49}, 3962 (1994);\\
      S.Z. Wu, F.O. Schumann, G.J. Mankey, and R.F. Willis, J. Vac. Sci. Technol. {\bf 14}, 3189 (1996).
   \bibitem{Binder1989} K. Binder and J.S. Wang, J. Stat. Phys. {\bf 55}, 87 (1989).
   \bibitem{Kadanoff1966} L.P. Kadanoff, Phys. {\bf 2}, 263 (1966).
\end{thebibliography}
\end{document}